\documentclass[preprint,prc,floatfix]{revtex4}
\usepackage{epsfig}
\usepackage{amssymb}

\begin{document}
\newcommand{\dbar}{$d$ \hspace*{-1.5ex}\raisebox{.7ex}{-} }
\bibliographystyle{apsrev}

\title{ $^4$He experiments can serve as a data base for determining the
three--nucleon force}
\author{ H.~M.~Hofmann }
\affiliation{Institut f{\"u}r Theoretische Physik III,
   University of Erlangen-N{\"u}rnberg, Staudtstra\ss{}e 7,
  D 91058 Erlangen, Germany}
\author{ G.~M.~Hale }
\affiliation{Theoretical Division, Los Alamos National Laboratory,
  Los Alamos, NM 87545, USA}

\begin{abstract}
We report on microscopic calculations for the $^4$He compound system
in the framework of the resonating group model employing realistic 
nucleon-nucleon and three nucleon forces. The resulting scattering phase shifts
are compared to those of a comprehensive $R$-matrix analysis of all data in this
system, which are available in numerical form. The agreement between calculation
and analysis is in most cases very good. Adding three-nucleon forces yields
in many cases large effects. For a few cases the new agreement is striking.
We relate some differences between calculation and analysis to specific data
and discuss neccessary experiments to clarify the situation. From the results 
we conclude that the data of the $^4$He system might be well suited to
determine the structure of the three-nucleon force.

\end{abstract}
\maketitle

\section*{Introduction}

For the nucleon-nucleon system the Nijmegen-group \cite{Nijmegen} has
developed a database of some 5600 data points for the proton-proton
and neutron-proton scattering. All modern
nucleon-nucleon potentials
have to be tested against this dataset and yield a $\chi{^2}$
per degree-of-freedom of the order of one before they are generally accepted.
For the three-nucleon force (TNF) the natural system are the three-nucleon
systems $^3$H and $^3$He. In these two cases, however, realistic 
nucleon-nucleon ($NN$) forces allow to describe the deuteron-nucleon scattering
data already quite well
\cite{KIEVSKY-ND} and the TNF yield only minor corrections for an almost
perfect reproduction of the data, except for the notorious $A_y$ problem
\cite{Ay}. But even to cure the $A_y$ problem only small changes of the
deuteron-nucleon $P$-wave phase shifts of the order of a degree are sufficient
to reproduce the low-energy data \cite{ND-PSA}. At higher energies, which we do 
not consider here, TNF effects are more pronounced; see \cite{ND-BO}.
Also, three-body breakup reactions might be a more sensitive test 
\cite{Gloeckle-breakupI,Gloeckle-breakupII,Gloeckle-breakup}. In this situation,
another system can be helpful to determine the TNF,
if such a system were accessible to scattering calculations employing
realistic two- and three-nucleon forces. Furthermore, the
available amount of data must be of the order of that of the $NN$ system.
At the moment computer power limits the scattering calculations in the 
many channel case to a mass number below 6.
For the three systems, $^4$H, $^4$He, and $^4$Li,
scattering calculations using realistic $NN$ forces exist \cite{ANT,HE4,Grenoble,
Lazauskas}.
Partially also TNF have been employed. The second criterion is only met by
the $^4$He system. The two-body scattering channels triton-proton,
$^3$He-neutron, and deuteron-deuteron allow for elastic scattering and various
reactions, with cross section measurements and also many polarization
observables existing. Due to well developed resonances \cite{TILLEY-A4} their
energy dependence is sometimes rather strong. About 5000 individual data are
the input for the ongoing $R$-matrix analysis \cite{GMH}, a number large enough
to allow for detailed comparison with calculations using various forces.

In this paper we report on microscopic calculations using realistic nuclear
forces, comparing the results for scattering phase shifts with those of the
$R$-matrix analysis on a partial-wave-by-partial-wave basis and then to a selected
set of data. This set of data is chosen to demonstrate the agreement to be reached by the analysis and the parameter free calculation, but also to indicate
the neccessity of new or better data in order to improve the $R$-matrix analysis
and to allow for a more detailed comparison with the calculated results.
The format of the paper follows to a large extent that of earlier work
\cite{HE4} using a rather old version of the Bonn potential \cite{Bonn} expanded in
terms of Gaussians \cite{BONNFIT}. We first discuss the essentials of the
refined resonating group model (RRGM) used here. Then we describe briefly the
$R$-matrix analysis, which is ongoing work from the evaluation \cite{TILLEY-A4},
and discuss changes since the last publication. A detailed comparison of
diagonal scattering phase shifts and reaction matrix elements from the analysis
with the calculated ones comprises the next section. The following section discusses
how well data can be 
reproduced by the $R$-matrix analysis and the calculation, and which conclusions
about the underlying interaction are possible.
Finally the quest for new or better data for selected experiments is
discussed, especially which kind of information might be drawn from them.
We conclude the paper by a discussion about the suitability of the dataset as
a measure for determining the structure of the TNF.

\section{RGM and model space}

We use the resonating group model \cite{ RRGM-VIEWEG, RRGM-TANG}
in its refined version  \cite{RRGM}
to compute the scattering in the $\rm ^4$He  system
using the Kohn-Hulth\'en variational principle \cite{KOHN}. The main
technical problem is the evaluation of the many-body matrix elements
in coordinate space. The restriction to a Gaussian basis for the
radial dependencies of the wave function allows for a fast and
efficient calculation of the individual matrix elements \cite{RRGM,
  RRGM-TANG}. However, to use these techniques the potentials must
also be given in configuration space 
in terms of Gaussians. In this work we use suitably
parametrized versions of the Argonne v18 (AV18) \cite{AV18} 
$NN$ potential, and the Urbana IX (UIX) \cite{AV8P-U8-U9},
and $V_3^*$ proposed in \cite{V3-SCHADOW} and used in \cite{BP}, $NNN$
potentials.

In the $^4$He system we use a model space with six two-fragment
channels, namely the $p$ - $^3$H, the $n$ - $^3$He, the $d$ - $^2$H,
the $d$ - $^2$H (S=0), the \dbar resonance, the \dbar - \dbar, and the
($pp$) - ($nn$) channels. The last three are an approximation to
the three- and four-body breakup channels that cannot in practice
be treated within the RRGM. The $\rm ^4$He is treated as four clusters
in the framework of the RRGM to allow for the required internal orbital
angular momenta of $\rm ^3$H, $\rm ^3$He or $\rm ^2$H.

For the scattering calculation we include all $S$-, $P$-, $D$-, and 
$F$-wave contributions to the $J^\pi = 0^+, 1^+, 2^+, 3^+, 4^+,
0^-, 1^-, 2^-, 3^- \text{, and }
4^-$ channels. From the $R$-matrix analysis these channels are known
to reproduce the low-energy experimental data.  The full wave function for these
channels contains over 100 different spin and orbital angular momentum
configurations, hence it is too complicated to be given in detail. 
The RRGM can be considered as a kind of variational calculation, hence, 
increasing the model space used usually improves the calculation. During the
work we increased the model spaces several times. Due to the amount of material 
to be shown, we present here only results obtained for the largest model space.
Some results obtained for smaller spaces are given in \cite{HLRB} and
\cite{SCAT-LE}.
Using a genetic
algorithm \cite{CWGEN} for AV18 and UIX together, and allowing for $S$-, $P$-,
and $D$-waves on all internal coordinates we found a triton binding energy of
-8.460 MeV for dimension 70.
This practically converged
result compares favorably with the numerically exact one of Nogga 
\cite{NOGGA-FAD} of -8.478 MeV. Since the Gaussian width parameters were
optimized for $NN$ and $NNN$ interaction together, the agreement for the
AV18 alone is only -7.57 MeV, compared to the exact one of -7.62 MeV \cite{NOGGA-FAD}.
Since isospin is a good approximation in light nuclei, we use for the $^3$H
and $^3$He the same internal configurations and the same width parameters;
however, the coefficients are (slightly) different due to the Coulomb force
(and to a minor extent isospin-breaking terms in the AV18 potential). Note,
that all calculations are done for physical channels, and not for channels
with good isospin.
For the deuteron we used 5 width parameters for the $S$-wave and 3 for the
$D$-wave, yielding -2.213 MeV, just 10 keV short of the experimental value of -2.2245 MeV.
The binding energies and relative thresholds for the various potentials are
given in table \ref{thres}. For $NN$ and $NNN$ together the experimental
binding energies and thresholds are very well reproduced.
The RGM can only deal with two-body channels. To mock up the break-up channels,
we allowed for configurations containing \dbar, ($pp$), and ($nn$).
Since these are unbound, any unrestricted variational calculation will produce
a wave function of infinite extent and zero energy. To have a finite extent,
we use the deuteron $S$-wave width parameters, and let only the coefficients
vary, to give the lowest possible (positive) energy. This procedure
yields as binding energies +0.818 MeV, +1.203 MeV, and +0.805
MeV, respectively. In the previous calculation \cite{HE4} these channels had 
some visible influence on the calculated phase shifts. In the present 
calculations they could be neglected, due to the much larger model spaces,
except for the binding energy of the $^4$He ground state, where they contribute
100 keV or 150 keV for AV18 alone and the TNF included. 

\begin{table}[h]
\centering
 \caption{\label{thres} Comparison of experimental and
 calculated total binding energies and thresholds relative to $^3{\rm H}-p$
 (in MeV) for the various potential models used}
 \vskip 0.2cm
 \begin{tabular}{c|c|c|c|c|c}
 \hline\noalign{\smallskip}
 potential & \multicolumn{3}{c|}{$E_{bin}$} & \multicolumn{2}{c}{$E_{thres}$} \\
     & $^3$H & $^3$He & $^2$H & $^3{\rm He}-n$  &  $ { d - d} $ \\
     \noalign{\smallskip}\hline\noalign{\smallskip}
     AV18    & -7.572& -6.857 & -2.213 & 0.715 & 3.145 \\
     AV18 + UIX & -8.460 & -7.713 & -2.213 & 0.747 & 4.033 \\
     AV18 + UIX +$V_3^*$ & -8.452 & -7.705 & -2.213 & 0.747 & 4.025 \\
     exp.    & -8.481 & -7.718 & -2.224 & 0.763 & 4.033\\
     \noalign{\smallskip}\hline
     \end{tabular}
     \end{table}

This representation of $^3$H/$^3$He , deuteron and the unbound $NN$ systems
form the model space of the $^4$He scattering system. We get for the
different $J^{\pi}$ values up to 10 physical channels, insufficient to find 
reasonable scattering results. 
So-called distortion or pseudo-inelastic
channels \cite{RRGM-TANG} without an asymptotic part
have to be added to improve the description
of the wave function within the interaction region. 
For this purpose all the configurations calculated for the physical channels
except one per channel can in principle be reused, keeping only those width parameters
which describe the internal region. In practice, however, this works only for a bound state
calculation. In scattering calculations, the numerical accuracy in manipulating large matrices
introduces some small amount of noise into the calculated results, e.g. phase shifts.
Therefore we omit from each physical channel two to five components, to avoid the noise completely.
This procedure reduces the binding energy of the ground state of $^4$He 
by typically 20 keV, relative to the full bound state calculation. 
In the following we will always give the energies from the
scattering calculation.

Recently  Fonseca \cite{NT-FONSECA} pointed out that states
having a negative parity $J_3^-$ in the three-nucleon fragments
increase the $n -^3$H cross section noteably. Contrary to the neutron-triton
system we found in the $^4$He system 
that the inclusion of such distortion states 
in the preliminary small model space calculations 
gave minor effects
compared to adding UIX. Therefore in the converged calculation we did
not allow for such states, in order to save computational resources, as we had 
anyhow to deal with sometimes more than a thousand channels.

\section{$R$-matrix analysis}
The Coulomb-corrected, charge-independent $R$-matrix analysis of the  
$^4$He system from which the various ``experimental" phase shifts are  
obtained in this paper is similar to the one described in Section 3  
of our previous publication \cite{HE4}. It uses  the approximate  
charge independence of nuclear forces to relate the parameters in  
charge-conjugate channels, while allowing simple corrections for the  
internal Coulomb effects.

The isospin $T=1$ parameters were taken from an analysis of  $p$ - $^3$He 
scattering data \cite {GMH} that gives a good description of all  
data at proton energies below 20 MeV. The $T=1$ eigenenergies $E_ 
\lambda^{T=1}$ are, however, shifted by the internal Coulomb energy  
difference $\Delta E_{C} = - 0.64$ MeV, and the $p$ - $^3$H and $n$ - $^3$He  
reduced-width amplitudes $\gamma_{c\lambda}^{T=1}$ are reduced by the  
isospin Clebsch-Gordan coefficient $1/ \sqrt 2$. The isospin $T=0$  
parameters are then varied to fit the experimental data for reactions  
among the two-fragment channels $p$ - $^3$H, $n$ - $^3$He, and   
$d$ - $^2$H, at energies corresponding to excitations in $^4$He  
below 29 MeV.  In this fit, the $T=0$ nucleon-trinucleon
reduced-width amplitudes are constrained by the isospin relation
$
\gamma ^{T=0}_{ n \, ^3{\rm He}} \; = - \gamma ^{T=0}_{ p \, ^3{\rm H}}
$,
and a small amount of internal Coulomb isospin mixing is introduced  
by allowing
$
\gamma ^{T=1}_{{ d d}} \neq 0,
$
which is neccessary to reproduce the differences between the two  
branches of the $d$ - $d$ reaction. The charge-independent constraints  
imposed on the parameters of this model might be too simple, and  
although the $R$-matrix results given here are not yet final, they  
represent the most comprehensive and detailed attempt to date to give  
a unified phenomenological description of the reactions in the $^4$He  
system. Note, that the $R$- and $S$-matrices are always represented in 
physical channels, and the isospin arguments are only used to reduce the
number of parameters in the $R$-matrix.

A summary of the channel configuration and data included for each  
reaction is given in table \ref {data}.  New data have been added in  
most of the reactions, including the neutron total cross sections of  
refs. \cite{Als-nie, NT-TOTAL-59, HAE83, KEI03}, the elastic  
scattering cross sections of \cite{ALF81}, and the $^3$H($p,n$)  
reaction cross-section measurements of \cite{GIB59, BRU99}.

\begin{table} [h]
\caption{Channel configuration (top) and data summary (bottom) for  
each reaction in the $^4$He system $R$-matrix analysis} \centering
\begin{tabular}{|c|c|c|}
\hline
Channel & $l_{\rm max}$ & $a_c$ (fm) \\ \hline $^3$H - $p$ &	3 &	4.9 \\ \hline
$^3$He - $n$ &	3 &	4.9 \\ \hline
$^2$H - $d$ &	3 &	7.0 \\
\end{tabular}\\
\begin{tabular}{|c|c|c|c|}
\hline
Reaction	& Energy range (MeV) & \# Observable types & \#
Data points
\\ \hline
$^3$H$(p,p)^3$H & $E_p=0-11$	&	~3	&
1382\\
$^3$H$(p,n)^3$He + inv.& $E_p=0-11$ &	~5	&
~856\\
$^3$He$(n,n)^3$He& $E_n=0-10$	&	~2	&
~397\\
$^2$H$(d,p)^3$H & $E_d=0-10$	&	~6	&
1666\\
$^2$H$(d,n)^3$He & $E_d=0-10$	&	~6	&
~921\\
$^2$H$(d,d)^2$H & $E_d=0-10$	&	~6	&
~399\\
\hline
&	totals:	&	28	&
5621\\
\hline
\end{tabular}
\label{data}
\end{table}

When the $S$-matrix is continued onto the complex energy
surface, near one of its poles it has the form
\begin{equation}
S = i \frac { \rho _{0} \rho ^{T}_{0} }{ E_{0} -E}
\end{equation}
where $E_{0} = E_{R} - i \Gamma \slash 2$ is
the complex pole energy and $\rho _{0}$ is the complex
residue amplitude. A procedure for obtaining $E_{0}$ and
$\rho _{0}$ from $R$-matrix parameters is given in \cite
{Ha4}. The expectation of the Breit-Wigner approximation is
that the sum of the partial widths is related to the imaginary part  
of the pole energy by
\begin{equation}
\rho_0^\dagger \rho_0=\Gamma.
\end{equation}
For the resonances in light systems such as $^4$He,
this is often not the case \cite {Ha4,Ko}. As explained in \cite  
{Ha4}, a parameter
characterizing the strength of an $S$-matrix pole,
\begin{equation}
{\cal S} = \frac { \rho ^{\dagger}_{0} \rho _{0} }{ \Gamma},
\end{equation}
in terms of the magnitude of its residue compared to its displacement  
from the real axis, can be quite different from unity for poles that  
do not show up as  strong resonances in the data.

\section{Phase shift comparisons}

Since the nuclear many-body forces are not yet well enough established, 
we cannot anticipate that a direct comparison between calculation and 
data leads to clear conclusions,
especially as the complicated, time consuming calculations do not allow for 
easy modifications of the potentials used.

It is enough that the matrix elements of one partial wave are not reproduced
well enough, to spoil any agreement between calculation and data. In this
situation the comprehensive $R$-matrix analysis, which takes into account all
physical channels simultaneously, connects elastic scattering data with
reactions and interpolates in energy, is an absolute must.
This interpolation is neccessary to study energy dependencies and the effect of,
even broad, resonances.

We can write each $S$-matrix element connecting channels $a$ and $b$ as
$\left < a, l_a, S_a | S^{J^{\pi}} | b, l_b, S_b \right > = \eta e^{2 i \delta}$,
where $\eta$ and $\delta$ depend on the channels $a$ and $b$, 
their orbital angular
momenta $l_a$ and $l_b$, their channel spins $S_a$ and $S_b$, and the total
angular momentum J and parity $\pi$.

In the following we will compare in most cases only diagonal scattering phase
shifts $\delta$ and only sometimes present results for the coupling strength
$\eta$, in order not to be swamped with too many data. In addition all
$S$-matrix elements to a given $J^{\pi}$-value have to obey unitarity; hence,
diagonal and non-diagonal matrix elements are always related.
 
\begin{figure}[h]
\centering
\includegraphics[width=8cm]{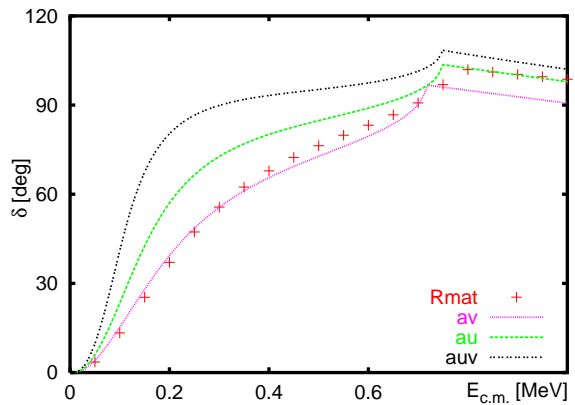}
\caption{(Color online) Low-energy triton-proton $0^+$ phase shifts calculated
using AV18 (av), AV18 and UIX (au), and additionally  $V_3^*$ (auv) compared to
$R$-matrix (Rmat) results.}
\label{tp-low}
\end{figure}

For most energies we have to deal with a coupled channels problem. Only below the $^3$He - $n$
threshold at 700 keV the triton-proton $0^+$, $0^-$, $3^+$,
and $4^-$ channels are single
channels indeed. In this energy range, due to the low energy, the $P$-wave channels are still dominated by the threshold
behavior, showing only small phase shifts. The $D$- and $F$-waves are even smaller,
below 0.1 degrees. The $0^+$ channel, however, is dominated by
the first exited state in $^4$He. The $R$-matrix analysis yields a rapidly increasing phase
shift, which crosses the 90-degree line 100 keV below the threshold, reaching about 105
degrees at threshold in a cusp-like manner, to slowly decrease afterwards, as shown in
fig. \ref{tp-low}.
The corresponding pole of the $S$-matrix is found much below, close to the triton-proton
threshold at a complex energy of 0.114 - i0.196 MeV.

The calculations started with the AV18 $NN$ potential alone. With increasing model space,
the $^4$He
binding energy converged at -24.090 MeV. All calculations showed qualitative agreement with the
$R$-matrix results; the smallest model space is above the $R$-matrix results, an increased one
well below - see fig.1  in ref.\cite{HLRB} - and the converged model space yields quantitative
agreement, as shown in fig.\ref{tp-low}. The corresponding $S$-matrix has in the low-energy region two
complex energy poles,
given in table \ref{0+-poles}. 
The lower one is close to the one found in the
$R$-matrix analysis.

For a well isolated resonance the modulus of the residue of the $S$-matrix 
at the pole position $E_R -i \Gamma /2$ 
in the single channel case is just $\Gamma$, dictated by unitarity.
As most of the resonances in $^4$He are broad and overlapping, this criterion is not
met in most cases, see the previous section and 
\cite{TILLEY-A4} for a more detailed discussion. In order to indicate the 
relevance of a complex energy pole of the $S$-matrix, we present in
table \ref{0+-poles} the pole positions found, together with the 
ratio {$\cal S$} of
the modulus of the residue and $\Gamma $ for an easier comparison.

\begin{table}
\centering
\caption{\label{0+-poles} {Pole positions of the $0^+$ $S$-matrix from the multi-level
$R$-matrix analysis compared to the results of 
   the various potential models used. All energies are given in MeV}}
\vskip 0.2cm
\begin{tabular}{c|c|c|c|c|c|c|c|c|c|c}
\hline\noalign{\smallskip}
\multicolumn{2}{c|}{$R$-matrix} & \multicolumn{3}{c|}{AV18} & \multicolumn{3}{c|}{AV18 + UIX}& \multicolumn{3}{c}{AV18 + UIX +$V_3^*$}  \\
 $E_R$ & $ \Gamma/2$ & $E_R$ & $ \Gamma/2$ & {$\cal S$} & $E_R$ & $ \Gamma/2$ & {$\cal S$} &
 $E_R$ & $ \Gamma/2$ & {$\cal S$} \\
	         \noalign{\smallskip}\hline\noalign{\smallskip}
 0.114 & 0.196& 0.198 & 0.256 & 1.40 & 0.105 & 0.129 & 1.54 & 0.091 & 0.077 & 1.52 \\
    &  & 0.497 & 2.114& 0.06 & 0.664 & 2.227 & 0.17 & 0.574 & 2.229& 0.14\\
 \noalign{\smallskip}\hline
 \end{tabular}
 \end{table}
The real part of the calculated poles closest to the threshold is by no means related
to that energy, where the corresponding phase shift crosses 90 degrees; see fig.\ref{tp-low}.
The second pole found in all the calculations has a small ratio {$\cal S$},
which means it 
cannot be a standard Breit-Wigner resonance as discussed in \cite{TILLEY-A4}. 
These findings indicate the overlapping of (many) resonances. Furthermore, reducing the
model space slightly changes the positions of all poles strongly, except for the lowest one.
Usually we find many more poles (with small strength) in the $S$-matrix
calculation than in the $R$-matrix analysis, whose influence on the physical
observables is expected to be weak. The next pole in the $R$-matrix analysis
is above 3 MeV. In this interval more poles of the $S$-matrix are found.

Since neglecting the Coulomb force shifts the lowest pole below
the $^3$H - $p$ threshold, so that it becomes a particle-stable state (see also
\cite{NHE-F} for various $NN$ forces and \cite{Lazauskas} for additional $NNN$
forces), we discuss
the situation in more detail. For AV18 alone this state is bound 
relative to the threshold by 62 keV, for AV18 and UIX by 112 keV. Therefore in 
these cases, the $0^+$ $^3$H - $p$ phase shift has to fall with energy and cannot reproduce
the $R$-matrix results.

\begin{figure}[h]
\hspace{-1.0cm}
\includegraphics[width=8cm]{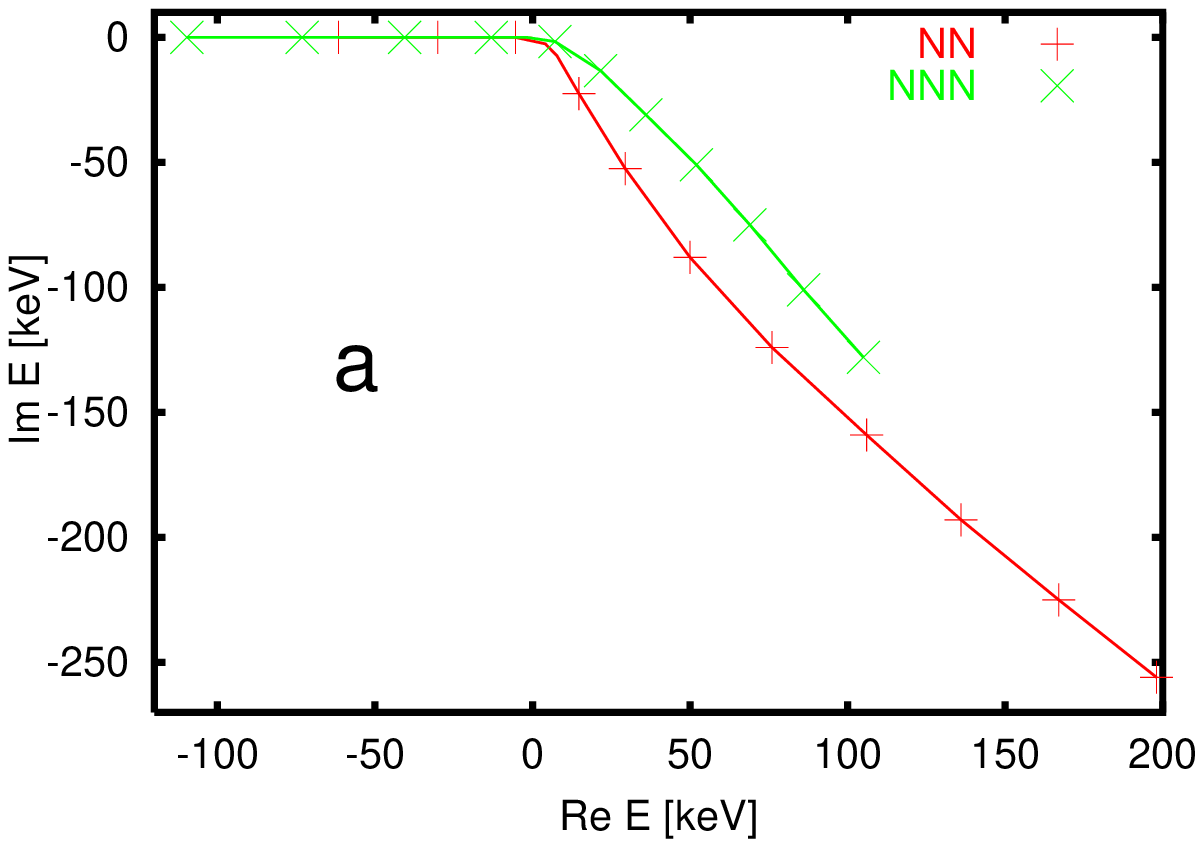}
\includegraphics[width=8cm]{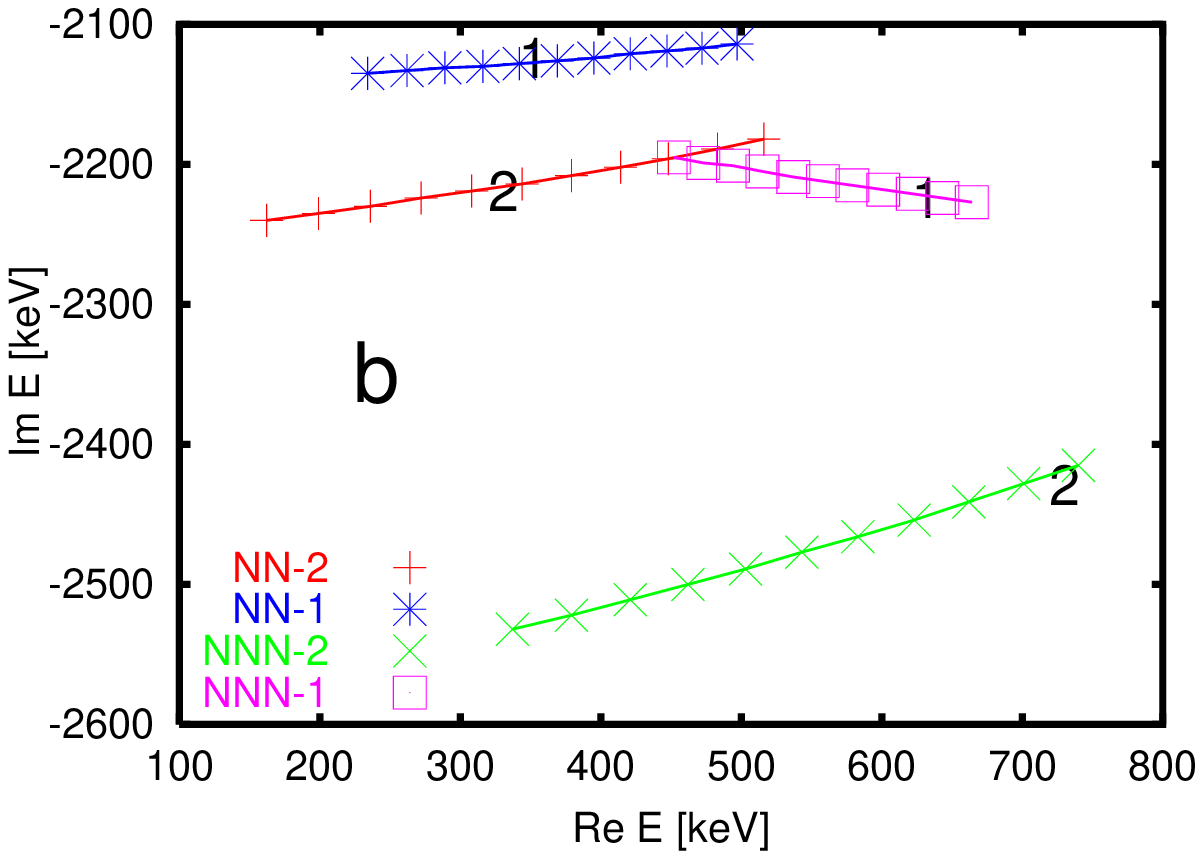}
\caption{(Color online) Complex energy pole positions of the $0^+$ $S$-matrix for a) the first
pole and b) the second pole. The ticks indicate the strength of the Coulomb
potential on a 0.1 grid, starting with zero most strongly
bound; for details, see text.}
\label{0p-poles}
\end{figure}

In order to sketch the intriguing situation we display the two $0^+$ $S$-matrix
poles closest to the $^3$H - $p$ threshold in fig.\ref{0p-poles} for the AV18 alone 
and the full interaction.

For the AV18 alone the second $0^+$ state starts just below -60 keV on the
real axis, becomes unbound slightly above a
strength of the Coulomb potential of $C=0.2$, then moves gradually into the 
complex energy plane up to a real part of the energy of about 200 keV. 
This pole is characterized by a residue that
is always larger than required by the unitarity condition on the real
axis by up to a factor of
1.9, close to the  $^3$H - $p$ threshold, falling to 1.4 for the full 
Coulomb interaction.

Employing now also the Urbana IX potential the situation is quite similar, 
as shown in fig. \ref{0p-poles}a. Since the second $0^+$ state is more strongly bound
without the Coulomb force, it becomes only unbound for a strength in excess
of $C=0.35$, then starting to move into the complex plane at a somewhat slower
pace, but with a similar pattern as for AV18. The residue is up to a factor of 
three larger than required by unitarity. Note, that all the resonant states
are well below the corresponding $^3$He - $n$ threshold. They are all on the
unphysical Riemann-sheet adjacent to the physical Riemann-sheet, with one
channel only open.

In table \ref{0+-poles} we find the next pole in energy still below the
 $^3$He - $n$ threshold for the full Coulomb strength. These states, however,
are not analytically connected to those found for no Coulomb interaction.
Except for a small splitting of 60 keV, due to the isospin breaking terms in
the AV18 potential the $^3$He - $n$ and $^3$H - $p$ thresholds coincide for
vanishing Coulomb interaction. Hence, the unphysical Riemann-sheet adjacent
to the physical sheet has two open channels. The corresponding pole positions
are labeled with a "2" in fig. \ref{0p-poles}b. With increasing Coulomb strength,
the real part of the pole position has to increase, due to the  $^3$He - $n$ 
threshold moving to higher energies and thus reducing the attraction.
We mark with "2" on the corresponding lines in fig. \ref{0p-poles}b, from which
strength onward the adjacent Riemann-sheet has only one open channel. This means
that the influence of these poles onto the observables on the real axis will
be reduced. Accordingly, we label pole position on the one-open-channel
Riemann-sheet with "1" and mark with "1" on the lines in fig. \ref{0p-poles}b,
from which strength on this Riemann-sheet is adjacent to the physical one.
For the AV18 alone this transition occurs around a strength of $C=0.45$,
yielding two poles with real energy positions close to the corresponding
$^3$He - $n$ threshold, but far in the complex plane with small residues.
Adding the UIX potential, the one-channel pole appears on the adjacent
Riemann-sheet just above $C=0.80$, whereas the two-channel pole disappears
only for a strength $C=0.97$. All the poles have very small residues, when they
are on the adjacent sheet to the physical one.

So altogether we are faced in both calculations with the situation of two poles,
one of which becomes a bound state increasing its residue when leaving the real axis,
and the other has a much smaller residue than expected from a standard
Breit-Wigner resonance. This is the only case that we encountered two
$S$-matrix poles of the same angular momentum and parity so close together in energy.
The behavior of the residues, one growing, one reduced when going into the
complex plane makes it very difficult to predict their respective effects
on the real axis, where they could be compared to experiment.
We mention in passing, that the exact position of the poles reacts very sensitive
on the numerical procedures of inverting the big matrices or how the regularized
Coulomb functions are expanded in terms of Gaussians.

Let us now discuss the behavior of the phase shifts, displayed in fig. \ref{tp-low}, in
more detail. We find the results for the AV18 $NN$ interaction alone in almost perfect 
agreement with the $R$-matrix analysis below the $^3$He - $n$ threshold, despite the
corresponding binding energy of $^4$He missing 4 MeV. When the $^3$He - $n$ channel 
opens, a bit too early (see table \ref{thres}),
the calculated triton-proton phase shift decreases slowly with energy.

Adding the UIX TNF force allows to reproduce the $^3$He - $n$ threshold much better, but below
threshold the calculated phase shifts overshoot the $R$-matrix results quite a bit. Above the
agreement is nice. The calculated binding energy of -28.294 MeV
is almost at the experimental value of -28.296 MeV.
With the additional TNF $V_3^*$ the phase shifts are much too positive,
the threshold is nicely reproduced and the ground state is overbound by 700 keV.
This interaction, tailored to resolve the $A_y$ problem in nucleon-deuteron scattering,
hence, acting in relative $P$-waves only, is
therefore not acceptable, and we will give no further results for this 
interaction. Furthermore, except for the above phase shifts
and close to the $^3$He - $n$ threshold,
the effects of this force are of minor importance. 
Directly above the $^3$He - $n$ threshold all potentials
yield a linear fall off with energy.

\begin{figure}[h]
\hspace{-1.0cm}
\includegraphics[width=8cm]{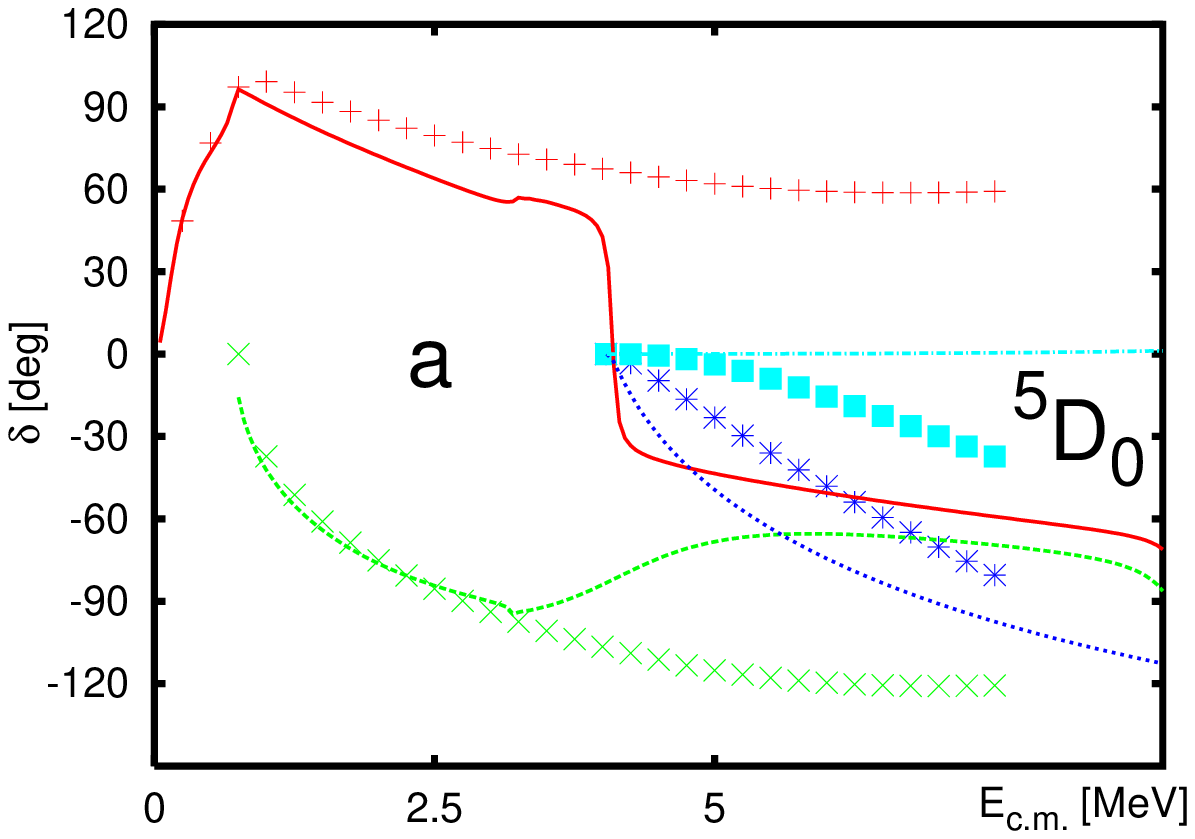}
\includegraphics[width=8cm]{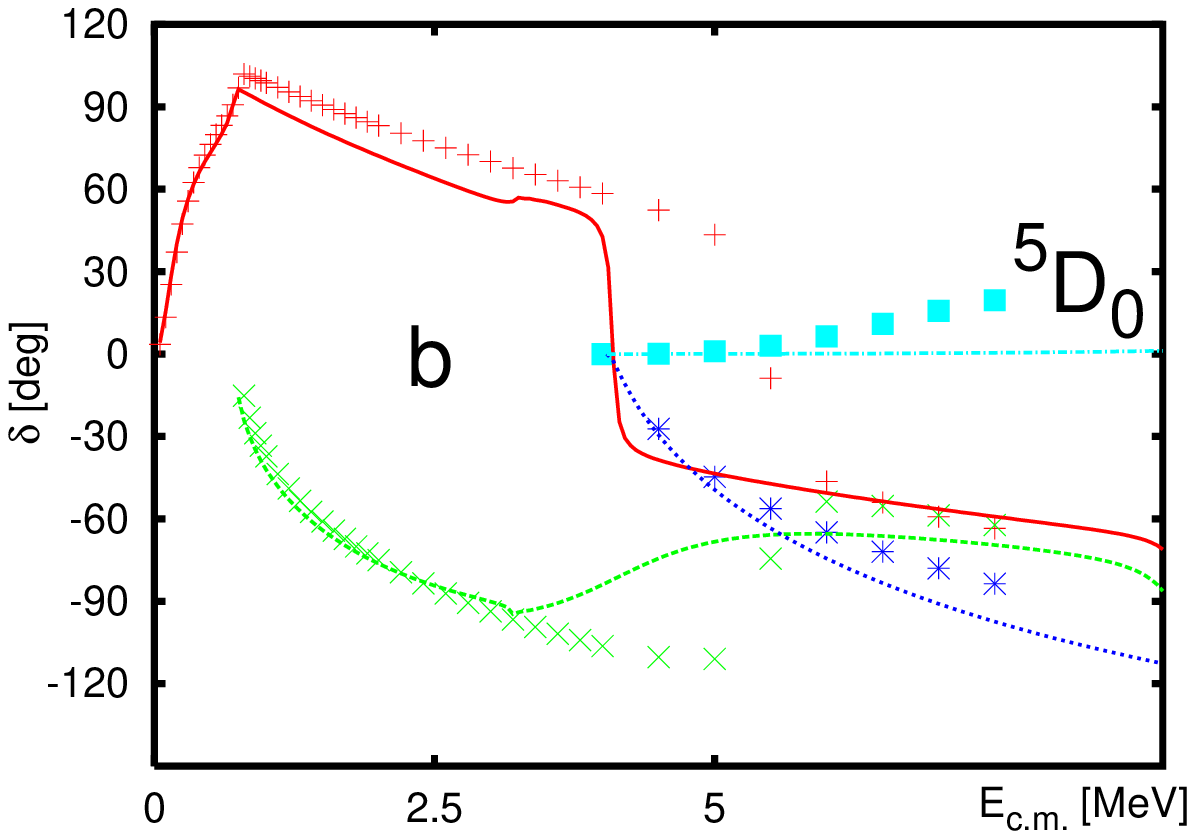}
\caption{(Color online) Elastic $0^+$ phase shifts for all physical two-fragment channels calculated for 
the AV18 potential alone. The RRGM $^1S_0$ phase shifts are displayed as full
lines (red) for the $t$ - $p$ ones, as dashed lines for the $^3$He - $n$ ones (green),
and as dotted lines for the $d$ - $d$ ones (blue). The corresponding $R$-matrix results
are given by + for $t$ - $p$, by x for $^3$He - $n$, and by $\ast$ for $d$ - $d$. We stick to this
coding in this subsection where ever possible. The $^5D_0$ $d$ - $d$
phase shifts are especially marked.
The RRGM calculated $d$ - $d$ phase
shifts are shifted by 0.9 MeV to the experimental threshold. (a) The data from
the $R$-matrix analysis are from 1997 as in \cite{HE4}. (b) The data are from the $R$-matrix 
analysis from 2003.}
\label{0p-av}
\end{figure}

Let us now discuss the higher energies. In \cite{HE4} the phase shifts resulting from a
version of the Bonn potential were discussed. The comparison of the diagonal phases from
the calculation and the $R$-matrix analysis in fig. 5 of ref. \cite{HE4} 
was not too convincing. Only an Argand plot
revealed that the diagonal $S$-matrix elements become small due to a very strong coupling
between the triton-proton and the $^3$He - $n$ channel to form a state of good isospin T = 0.
Therefore the influence of these matrix elements on physical observables is weak.

In fig. \ref{0p-av} we display the calculation for the AV18 potential alone in comparison with
the results from the $R$-matrix analysis used in 1997 \cite{HE4} and a recent
one from 2003 \cite{GMH}.
The differences in the data input into these two analysis are discussed above.
The obvious change in the data between the analysis from 1997 and 2003 is in the triton-proton
and  $^3$He - $n$ phases. Above 5 MeV they come close to each other as already predicted
in the old calculation
\cite{HE4} and differ no more by 180 degrees as before. The $^1S_0$ $d$ - $d$
phases became more repulsive close to the threshold.
The other change is the sign flip
in the $^5D_0$ $d$ - $d$ channel. But these phases are small anyhow.
The RRGM calculated results follow much more closely both $R$-matrix results up to about 3 MeV,
compared to \cite{HE4}. At the calculated $d$ - $d$ threshold there occurs a change in form.
(Note that the calculated $d$ - $d$ phase shifts have been shifted in energy to the experimental
threshold for an easier comparison. The tiny kink in the triton-proton phase shifts occurs
at the calculated $d$ - $d$ threshold.) The energetic position of the calculated results are
obviously wrong, due to the calculated $d$ - $d$ threshold being much too low; see table \ref{thres}. The qualitative
behavior is encouraging. Since the only qualitative change in the $R$-matrix results of
both analyses is in the $0^+$ phase shifts and the quantitative changes in all other
partial waves are too small to be clearly identified in the figures, we present in the following
only results from the 2003 $R$-matrix analysis.

\begin{figure}[h]
\centering
\includegraphics[width=8cm]{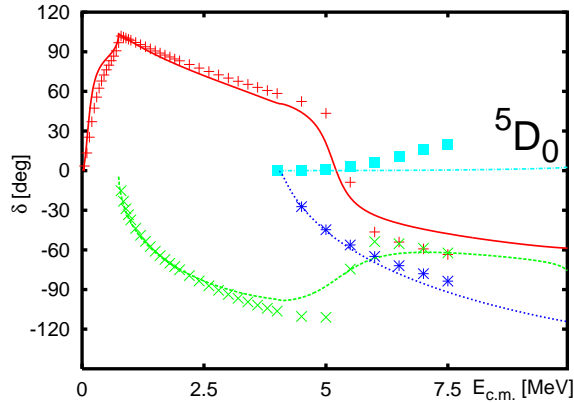}
\caption{(Color online) Elastic $0^+$ phase shifts for all physical two-fragment channels calculated for 
the AV18 potential together with the UIX TNF. The coding of the lines and 
symbols is the same as in fig. \ref{0p-av}.}
\label{0p-au}
\end{figure}

In fig. \ref{0p-au} we compare all physical $0^+$ elastic scattering phase shifts
calculated for AV18 together with UIX TNF with the recent $R$-matrix analysis.
The agreement between this zero-parameter calculation and the analysis is remarkably good. 
As discussed above, the
calculated low energy triton-proton phases are a bit too high, but from the  $^3$He - $n$ threshold 
up to the $d$ - $d$ threshold RRGM calculation and $R$-matrix
analysis agree perfectly, then comes the rapid change
in energy in the $^3$H - $p$ and $^3$He - $n$ phases, settling at similar values around 7 MeV.
The $^1S_0$ $d$ - $d$ phase shifts are very well reproduced over the analyzed
interval.
Obviously these results do not leave too much room for modifications of the underlying
$NN$ and $NNN$ potentials acting in the relative $S$-waves.
The small values of the $^5D_0$ phases are not reached by the calculation.

\begin{figure}[h]
\centering
\includegraphics[width=8cm]{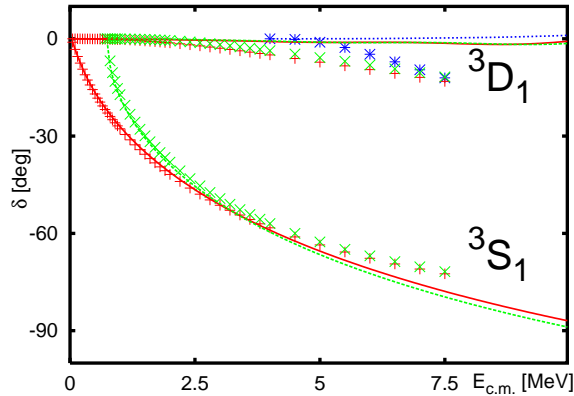}
\caption{(Color online) Same as fig. \ref{0p-av} but for the elastic $1^+$ phase shifts
calculated for the AV18 potential together with the UIX TNF.}
\label{1p-au}
\end{figure}

Next we discuss the $J^{\pi} = 1^+$ partial waves.
There the $R$-matrix analysis finds only
one high lying resonance; see ref. \cite{TILLEY-A4}. In fig. \ref{1p-au} we compare the full
calculation with the analysis. All phase shifts from the $R$-matrix
are negative. The $D$-wave ones are close to zero and do not reach -15 degrees,
and the $d$ - $d$ ones are
a bit more negative than in the earlier analysis; see fig.7 of ref. \cite{HE4}.
Both RRGM $S$-wave phase shifts agree nicely with the $R$-matrix
results up to 6 MeV, above which the RRGM ones are a bit more repulsive than those from the analysis.
The calculated $D$-wave phases are practically zero, the $d$ - $d$ ones slightly positive, the others
negative, but do not reach the $R$-matrix values, which are small anyhow.
Omitting these small phase shifts yields changes in the polarizations of
typically 0.01, which is of the order of the experimental error bars for
e. g. triton-proton scattering, but of the full data for deuteron-deuteron
scattering, see section IV.

\begin{figure}[h]
\centering
\includegraphics[width=8cm]{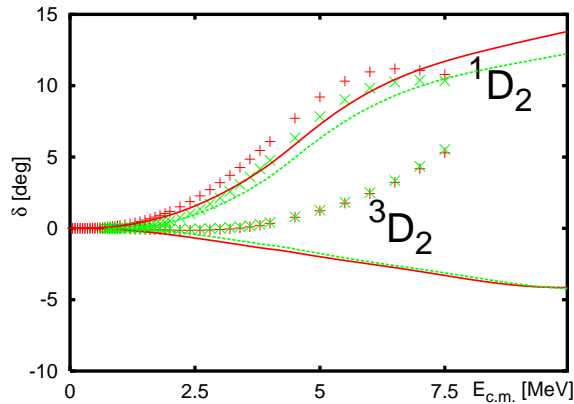}
\caption{(Color online) Comparison of the $2^+$ $t$ - $p$ and $^3$He - $n$ phase shifts calculated for 
the AV18 potential together with the UIX TNF. }
\label{2p-au}
\end{figure}

The $2^+$ partial wave contains the most coupled channels. In the $R$-matrix analysis all 
possible $D$-waves and the $^5S_2$ $d$ - $d$ channel were taken into account. In the RRGM calculation
the same physical channels and additionally, more than a thousand distortion channels are
considered. Since all the $D$-wave phase shifts are small already, the $^5G_2$ $d$ - $d$ channel
is neglected. In fig. \ref{2p-au} we compare the $R$-matrix [ 3 + 1 ] phase shifts with the 
calculated ones. All $R$-matrix phases appear to be 
positive, the $^1D_2$ just passing $10^\circ$ with
the $t$ - $p$ ones always a bit larger than the $^3$He - $n$ ones. The $^3D_2$ phases
start out with tiny negative values up to -0.1 degree, before turning positive
around 3 MeV, and barely reach
$5^\circ$. All phases are slightly more positive than in the previous analysis \cite{HE4}. 
The RRGM calculated $^1D_2$ phase shifts agree nicely with the analysis, a bit below at
low energies, passing the $R$-matrix results at the end of the analyzed interval, but keep
growing with energy, whereas the $R$-matrix ones show signs of decreasing. The $^3D_2$ 
calculated phases are also small and yield very similar values for $t$ - $p$ and  $^3$He - n,
but they disagree in sign with the $R$-matrix ones at higher energies.
Since an earlier $R$-matrix analysis gave also a negative sign for
these phase shifts, we were quite concerned, and tried to identify which of the 
additional data in the analysis might have introduced this sign flip. We could not find
any polarization observable sensitive to the sign flip, and will come back to this point
when discussing the comparison with data later on.

\begin{figure}[h]
\centering
\includegraphics[width=8cm]{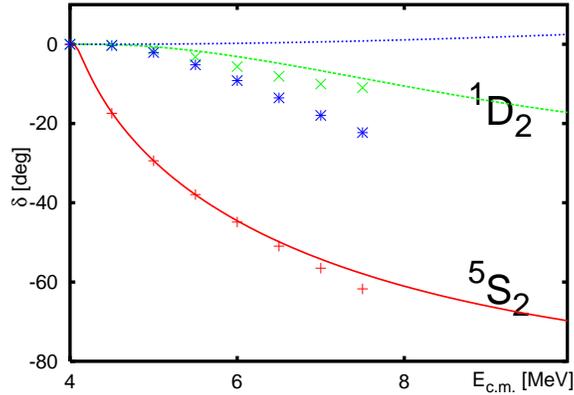}
\caption{(Color online) Comparison of the $2^+$ $d$ - $d$  phase shifts calculated for 
the AV18 potential together with the UIX TNF. The $^5S_2$ are shown as full
line (red) and +, the $^1D_2$ as dashed line (green) and x, and the $^5D_2$
as dotted line (blue) and $\ast$.}
\label{2p-dd-au}
\end{figure}

In fig. \ref{2p-dd-au} we display the corresponding
$d$ - $d$ phase shifts. The $^5S_2$ phases agree almost perfectly,
fall off rapidly with energy, but not as much as in ref. \cite{HE4}. The $D$-wave phase shifts
are small. The values of both channels from the $R$-matrix turned negative;
those from the RRGM calculation
are essentially unchanged. The $^1D_2$ phase shifts agree very well between calculation 
and analysis; the $^5D_2$ do not, indicating a rather large J-splitting in the
$R$-matrix analysis, and essentially none in the RRGM calculation.

\begin{figure}[h]
\hspace{-1.0cm}
\includegraphics[width=8cm]{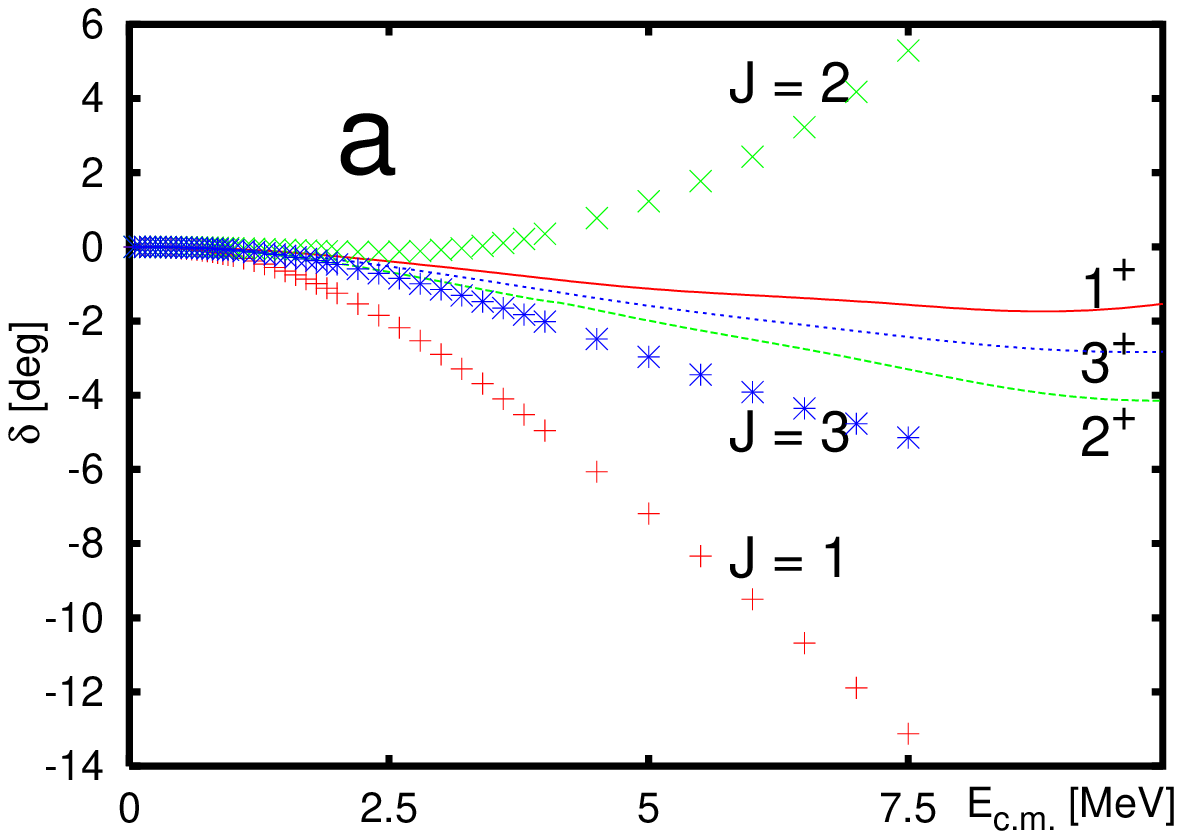}
\includegraphics[width=8cm]{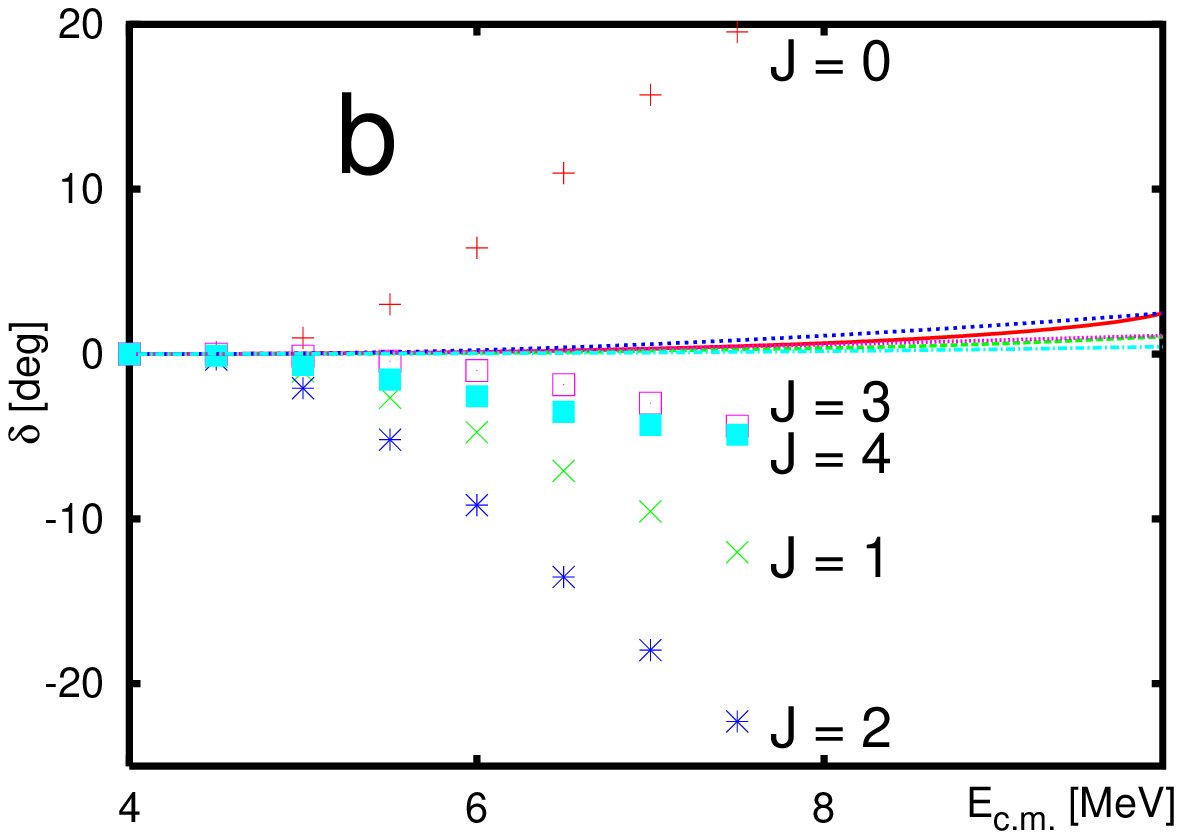}
\caption{(Color online) Comparison of the $D$-wave phase shifts for all possible total angular
momentum values J for (a) triton-proton channels and (b) the deuteron-deuteron 
channels. Due to the smallness of the RRGM $d$ - $d$ phase shifts, we do not 
label them.}
\label{D-wave-splitting}
\end{figure}

For the $3^+$ and $4^+$ partial waves the $R$-matrix analysis finds small negative values
for all $D$-wave phase shifts up to $-5^\circ$ for all physical channels.
The RRGM results are even smaller. These small phase shifts have in most cases
negligible effects on the observables, and are therefore not shown in a
separate figure.
Collecting, however, all $^3D_J$ triton-proton or $^3$He-neutron
or all $^5D_J$ deuteron-deuteron phase shifts from figs. \ref{0p-au}, \ref{1p-au},
\ref{2p-au}, and \ref{2p-dd-au} together with the results just mentioned,
we find considerable J-splitting from the $R$-matrix analysis and essentially none
in the RRGM calculation. This is shown in fig. \ref{D-wave-splitting}, where we display the
triton-proton triplet $D$-wave phase shifts -- the $^3$He - $n$ ones are similar --
and the deuteron-deuteron quintet $D$-wave phase shifts. Note that at the same
energy above the thresholds, the $R$-matrix analysis yields $d$ - $d$ phase shifts
that are much larger, and show a much larger splitting than the triton - proton
ones. it is remarkable that  all the $d$ - $d$ phase shifts are negative,
except for J = 0.

Let us summarize the results for the positive parities: Employing the AV18 and UIX
potentials allows to reproduce all $S$-wave phase shifts for all fragmentations almost
perfectly. Except for the $0^+$ partial wave these are dominated by Pauli repulsion
and, hence, negative. The singlet $D$-wave phase shifts agree nicely between analysis and
calculation. For the $^3D_J$ [ 3 + 1 ] phase shifts, the $R$-matrix yields large J-splitting,
with $^3D_1$ and $^3D_3$ negative, the other one positive. In a triton-proton
optical model, such a splitting could be caused
by a rather strong tensor force, but the actual values cannot be explained in a
perturbative treatment.

\begin{figure}[h]
\centering
\includegraphics[width=8cm]{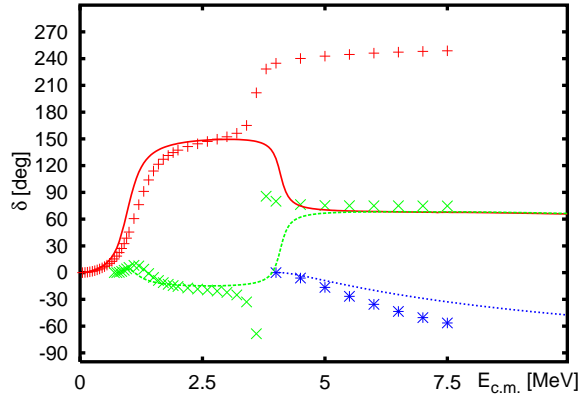}
\caption{(Color online) Same as fig. \ref{0p-av} but for the $0^-$ phase shifts. Above 4 MeV
we added 180 degrees to the $R$-matrix  $^3$He - $n$ phase shifts for an easier comparison
with the RRGM results.}
\label{0m-au}
\end{figure}

The corresponding $^5D_J$ $d$ - $d$ phase shifts cannot be accounted for by
deuteron-deuteron spin-orbit and tensor optical potentials, mainly due to the
J = $0^+$ phase shift being large and positive, and the J = $3^+$ and J = $4^+$
ones being small and negative. The RRGM calculation yields small negative values for the [ 3 + 1 ]
phase shifts (see fig.\ref{D-wave-splitting}), and a splitting of the
order of one degree, and for the $d$ - $d$ phase shifts, even smaller values, but
positive, with a similar splitting at comparable energies. These differences
in the values of the phase shifts and their splittings are the main
qualitative difference between $R$-matrix analysis and microscopic calculation
for diagonal $S$-matrix elements.
We will discuss this situation together with selected data at the end of the
paper.

Let us now consider the negative-parity partial waves. The compilation \cite{TILLEY-A4}
shows three $0^-$ resonances. Two of them can be easily read off from the rapid
change with energy of the $t$ - $p$ and $^3$He - $n$ phase shifts in fig. \ref{0m-au}.
The third one of $d$ - $d$ structure is not obvious. Except for the regions where the
[ 3 + 1 ] phases vary rapidly, the $t$ - $p$ and $^3$He - $n$ phase shifts differ
essentially by irrelevant multiples of $180^\circ$. In the region of the first resonance,
$R$-matrix and RRGM results agree nicely, but the position of the resonance seems to
be a bit lower in the calculation. In the region of the second resonance, 
which seems to be somewhat higher in the calculation, the
behavior of the $R$-matrix and RRGM phase shifts is completely opposite. Due to good
isospin T = 1, this resonance leads to a strong coupling of the two [ 3 + 1 ] channels,
with the coupling matrix element close to the unitary limit of one, leaving only
a small value for the diagonal $S$-matrix element. The situation we find here is
very close to that discussed previously \cite{HE4} and is explained by the 
Argand plots displayed in figs. 11 and 12 of ref. \cite{HE4}, by showing the two
$S$-matrices pass on different sides of the origin. Compared to the previous 
analysis and the calculation using the Bonn potential, the [ 3 + 1 ] phase shifts
are now much better reproduced, the $d$ - $d$ ones are a bit underestimated;
compare fig. \ref{0m-au} with fig. 10 of \cite{HE4}. Since no more recent
Bonn potential is given in r-space, it could not be tested, if these
differences are due to deficiencies of the older potential or to the much
increased model spaces.

\begin{figure}[h]
\centering
\includegraphics[width=8cm]{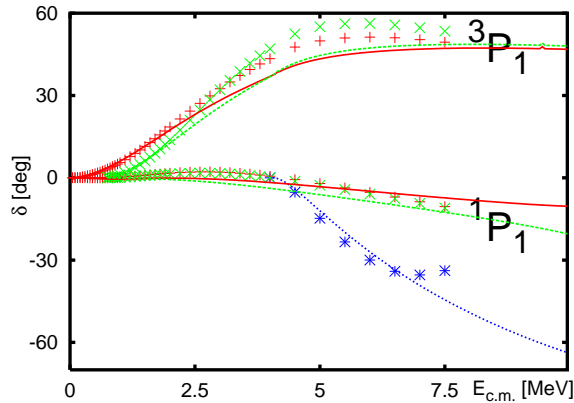}
\caption{(Color online) Same as fig. \ref{1p-au} but for the $1^-$ phase shifts.}
\label{1m-au}
\end{figure}

The $1^-$ phase shifts are displayed in fig. \ref{1m-au}. The extracted level structure
is quite rich \cite{TILLEY-A4}. The [ 3 + 1 ] triplet phase shifts are positive, all
others negative. The RRGM results agree favorably with the new $R$-matrix analysis. The
difference in the [ 3 + 1 ] triplet phase shifts is much smaller than for the Bonn
potential, as can be seen in fig. 13 of ref. \cite{HE4}. 
The $^3$He - $n$ triplet phase shifts
cross over the triton-proton ones at a higher energy and stay closer together
in the calculation. The $^1P_1$ phases show a small splitting that is not found
in the $R$-matrix analysis. Close to threshold, the $d$ - $d$ phase shifts are very
well reproduced. The difference at the higher energies in the 
$d$ - $d$ phase shifts could be due to end-of-data effects in the analysis.

\begin{figure}[h]
\hspace{-2.0cm}
\includegraphics[width=8cm]{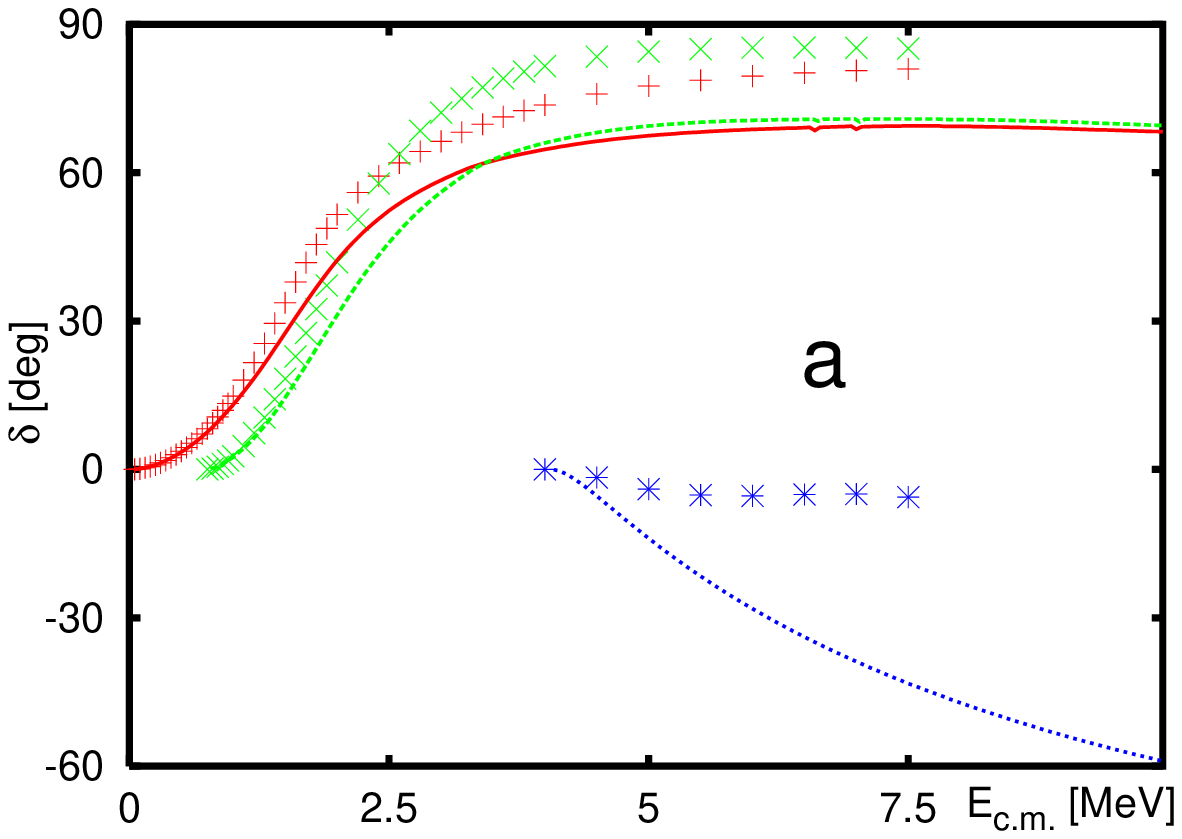}
\includegraphics[width=8cm]{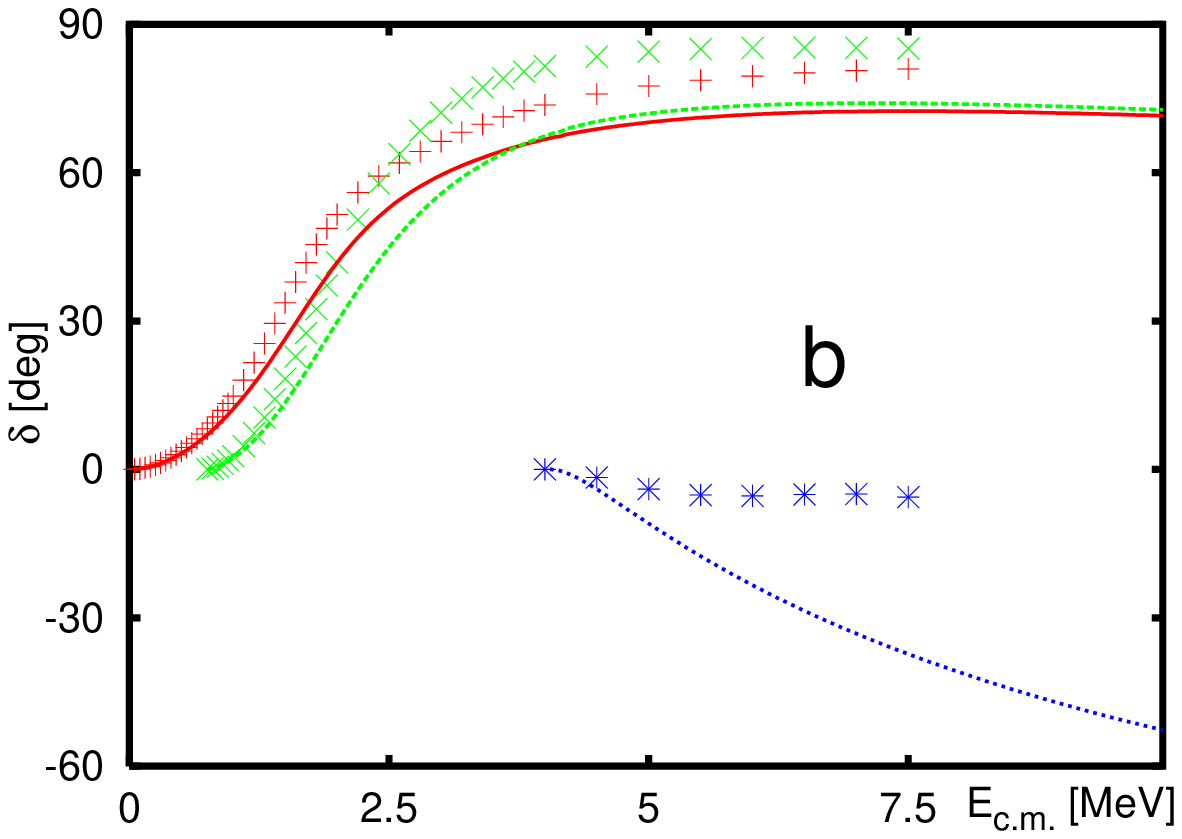}
\caption{(Color online) Comparison of the $2^-$ $P$-wave phase shifts. (a) Calculation for the
AV18 alone, no coupling to $F$-wave allowed. (b) Calculation for AV18 and UIX
including coupling to $F$-waves.}
\label{2m-auF}
\end{figure}

Since the $^3P_2$ phase shifts were the main culprit in \cite{HE4} for missing most
of the experimental values, we compare in fig. \ref{2m-auF} the new 
calculation for the $NN$ interaction alone and together with the TNF with 
the new analysis. The $R$-matrix values hardly changed; only the $d$ - $d$ phases 
became a bit less repulsive. Compared to the Bonn potential used in ref.
\cite{HE4}, the RRGM
results for the AV18 already are much more attractive, gaining 15 to 20 degrees at the
higher energies. Also the moduli of the $S$-matrix elements are much closer to the
$R$-matrix results than before.
Adding the UIX TNF yields a small amount of further attraction.
The $R$-matrix analysis allows also for $F$-waves. These can be coupled to $2^-, 3^-,$ and
$4^-$ partial waves. The resulting phase shifts turned out to be of the order of a
few degrees in the energy range considered. Therefore, and due to a lack of computing
power, these partial waves were not calculated in \cite{HE4}. During the comparison
with data it turned out that all the $^2$H($d$,nucleon) analyzing powers depend
sensitively on these small contributions and a description of the experimental
situation is only possible if these partial waves are taken into account. Therefore
every effort was made to calculate these. In fig. \ref{2m-auF}b we display the results
of the calculations allowing also for the UIX TNF and the $F$-waves.
The combined effect of coupling to $F$-waves and adding UIX TNF 
is additional attraction of the order of a few degrees.

\begin{figure}[h]
\hspace{-2.0cm}
\includegraphics[width=8cm]{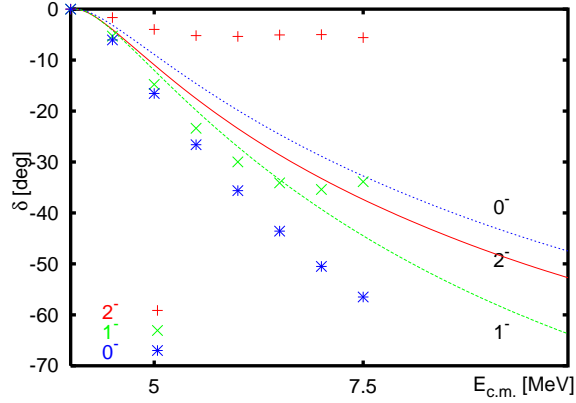}
\caption{(Color online) Comparison of the $d - d$  $P$-wave phase shifts.
The calculations are for
AV18 and the UIX potentials. The $^3P_2$ results are shown as full line (red)
and +, the $^3P_1$ as dashed line (green) and x, and the $^3P_0$ as dotted
line (blue) and $\ast$.}
\label{dd_P}
\end{figure}

As for the $1^-$ channels, the $^3$He - $n$ phase shifts cross over the $^3$H - p
ones at a higher energy, and have less splitting in the calculation than in the
$R$-matrix analysis. The $d$ - $d$ phase shifts from the analysis are close to zero,
whereas the calculation yields similar values as for the $1^-$ channel.
Altogether, the calculated [3 + 1] phase shifts are close to $R$-matrix ones,
whereas the calculated $d$ - $d$ phase shifts do not reach the J-splitting found
in the analysis.
To make the situation more transparent, we compare in fig. \ref{dd_P} the 
calculated deuteron-deuteron $P$-wave phase shifts with those from the analysis.

As seen in fig. \ref{dd_P}, the $^3P_1$ phase shifts agree nicely between 
calculation and analysis, and also the amount of splitting between the $^3P_1$ 
and $^3P_0$ phases is similar, but opposite in direction. The order of the
different J-values is permuted. The $^3P_2$ phases vary widely between 
calculation and analysis, but due to the small phase shifts in the analysis
we do not expect large effects omitting them altogether. Considering the three
J-values together, we find a reasonably large splitting in the analysis, hence,
also reasonably large polarization data. From the small J-splitting of the
calculated phase shifts, we expect small polarization data in d-d scattering.
If the analysis and the direct calculation are to reproduce the same d-d
data, the larger splitting in the $R$-matrix analysis has to be compensated by the
results from higher partial waves, due to interference.
We will come back to this point
when discussing elastic deuteron-deuteron observables later in the paper.

\begin{figure}[h]
\hspace{-2.0cm}
\includegraphics[width=8cm]{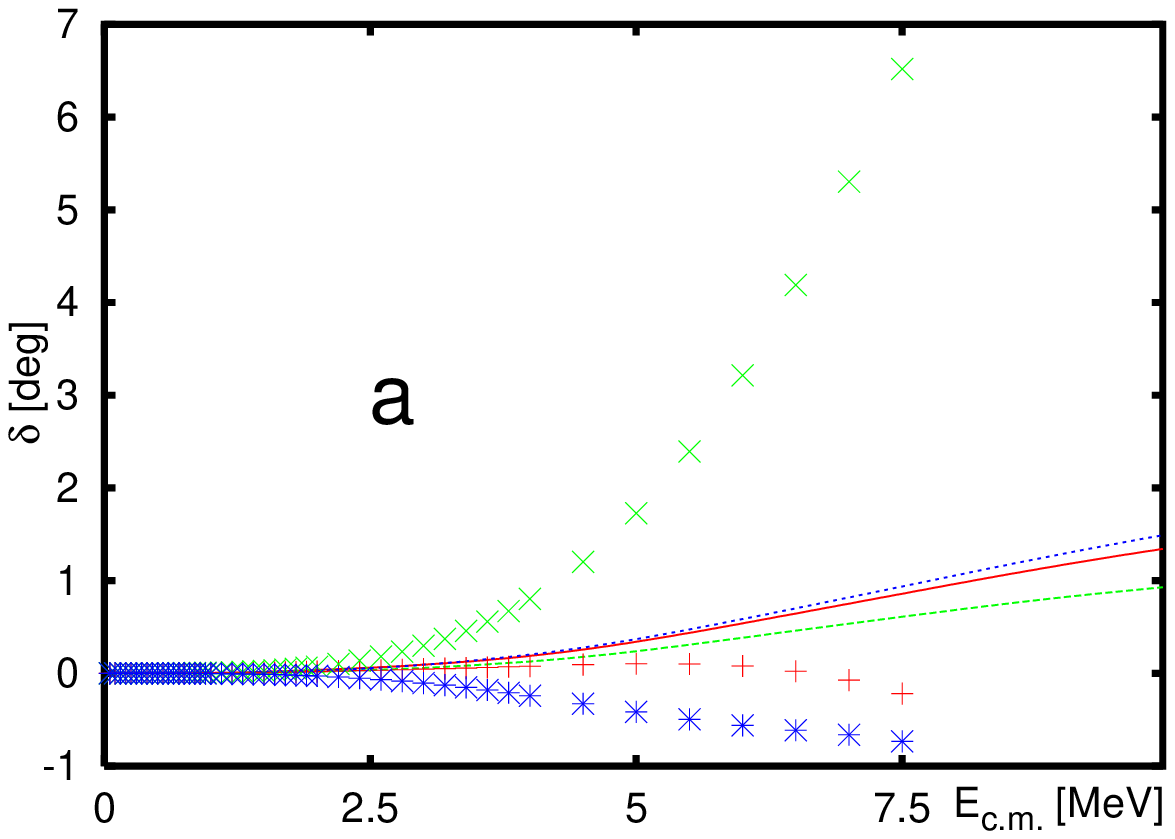}
\includegraphics[width=8cm]{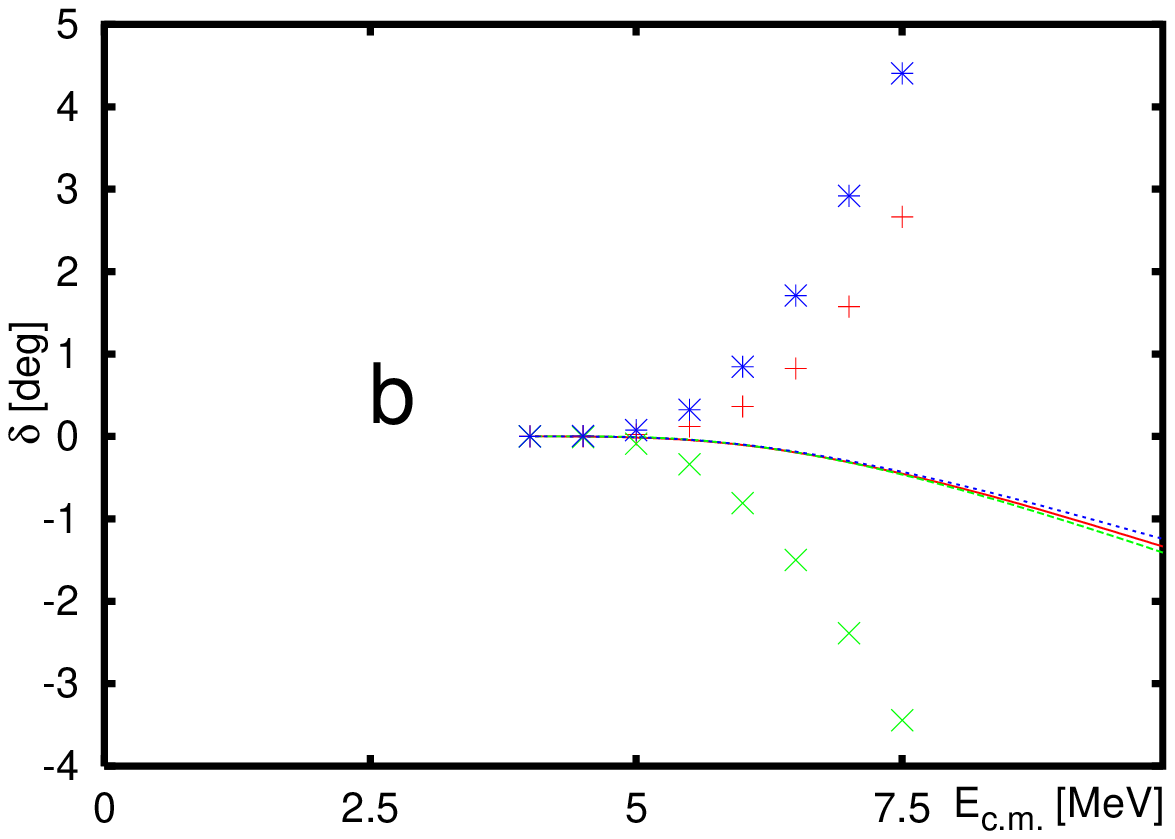}
\caption{(Color online) Comparison of the triplet $F$-wave phase shifts (a) for the
triton-proton channels and (b) the deuteron-deuteron channels. The calculations are for
AV18 and the UIX potentials. The $^3F_2$ results are shown as full line (red)
and +, the $^3F_3$ as dashed line (green) and x, and the $^3F_4$ as dotted
line (blue) and $\ast$.}
\label{F-wave}
\end{figure}

In fig. \ref{F-wave} we display the $F$-wave phase shifts for the $^3$H - $p$
and $d$ - $d$ channels. Since the results for the $^3$He - $n$ channels are almost identical,
we do not show them. The triton-proton $R$-matrix results are
small, the $^3F_3$ phase shift being reasonably large and positive up to $7^\circ$,
whereas the others are negative by less then 1 degree. The opposite sign of the  $^3F_3$
phases could be due to a strong tensor force; however, 
the size of the $^3F_2$ and $^3F_4$
phases does not support this conjecture. The RRGM calculation yields
essentially no J-splitting with all phase shifts slightly positive, up to one degree.
We note in passing that the singlet $F$-wave phase shifts have the opposite 
sign to the $^3F_3$ ones, and about half the strength in the analysis and the calculation.

For the deuteron-deuteron channels the $R$-matrix analysis yields a reasonably large
J-splitting with again the $^3F_3$ phase shift having the opposite sign of the rest.
The RRGM calculation yields again essentially no splitting; all the phases are small and
negative. This qualitative difference between $R$-matrix analysis and RRGM
calculation seems much less important than for the $D$-waves
because the values are much smaller; see, however, the discussion of the
deuteron-deuteron analyzing powers below.

\section{Direct comparison with experiments }

In order to allow the Reader a direct comparison with the previous results obtained
for the Bonn potential \cite{HE4}, we present figures for all the elastic scatterings
and reactions presented there. We always use the results of the recent $R$-matrix 
analysis \cite{GMH} and compare to the most complete calculations,
i.e. using the
AV18 $NN$ potential alone and together with the UIX TNF potential and taking into 
account all $S$-, $P$-, $D$-, and $F$-wave matrix elements.

\begin{figure}[h]
\centering
\includegraphics[width=8cm]{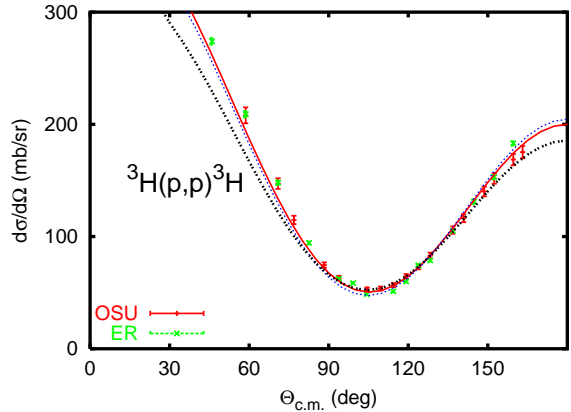}
\caption{(Color online) Differential elastic proton-triton cross section calculated at 3.1 MeV 
E$_{cm}$. The $R$-matrix results are shown as full line (red), the results from 
AV18 alone as thin dotted line (blue), and those for AV18 together with UIX as
thick dotted line (black). We stick to this coding in the following where possible.
The data are from Ohio State University (OSU) group \cite{Ohio} and from the
Erlangen group \cite{ER-tp}at 3.11 MeV.}
\label{tptp-x}
\end{figure}

So contrary to \cite{HE4}, the $R$-matrix analysis and the RRGM calculation now consider
exactly the same channels.
When the inclusion of $F$-waves yields substantially different results, we also 
present those without $F$-waves.
The RRGM calculation
is done in 50 keV steps in the center-of-mass, starting from the $^3$H - $p$ threshold.
This yields small deviations in energy from the experimental numbers, but the errors
introduced by this procedure should be well within the size of the points used.
The $R$-matrix analysis uses relativistic kinematics, the experimental
threshold energies, and the correct energies of the data for the fit. Here we present
results calculated for a varying energy grid, which uses values quite
close to the experimental numbers, but calculated with
non-relativistic kinematics. These small differences 
do not play any significant role for the examples given in the following.  We 
display first the data together with analysis and calculation as shown
in \cite{HE4}, sometimes adding new data at the same energy. 
Then we discuss a few data sets
that we consider critical to a further new analysis or to conclusions about the
effects of TNF forces.

We present the various reactions in the order of the corresponding thresholds,
starting with triton-proton elastic scattering. Around 4 MeV proton energy,
differential
cross section and analyzing-power measurements exist. In fig. \ref{tptp-x} we
compare the cross section data with the $R$-matrix analysis and the direct
calculation.
We see the data covering an angle range from about $45^\circ$ to $160^\circ$ in the center-of-mass system, with the Erlangen data \cite{ER-tp}
having the smaller errors, but disagreeing with the OSU data \cite{Ohio} at 
backward angles. 
The RRGM calculation is at an energy of 3.1 MeV in the center of mass,
to be in agreement with the energy of the proton analyzing power;
the $R$-matrix is at 3.0 MeV. These energy differences are too small to show any 
effect in the figures.
The $R$-matrix reproduces the data quite nicely in general, and 
falls in between the data at the backward angles. For the AV18 potential alone,
the RRGM yields results much better than for the Bonn potential
\cite{HE4}, now almost agreeing with the $R$-matrix results, being slightly
below at forward angles and slightly above at backward angles.
Adding the UIX TNF destroys the good agreement. Now the calculation is well
below $R$-matrix and data at forward angles and also on the lower side for
backward angles, but slightly above the data and analysis in the minimum region.

\begin{figure}[h]
\centering
\includegraphics[width=8cm]{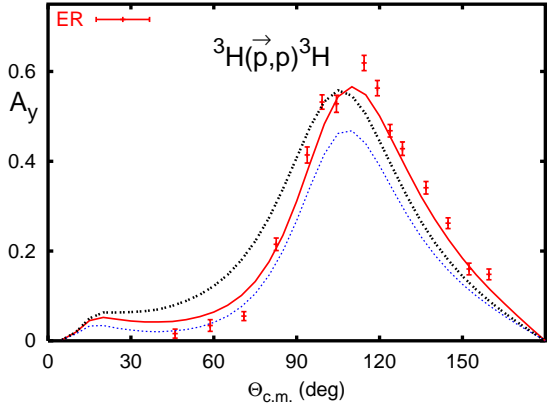}
\caption{(Color online) Proton analyzing power of the elastic scattering $^3$H($p,p$)$^3$H
calculated at 3.1 MeV E$_{cm}$. The meaning of the lines is as in fig. \ref{tptp-x}. The data are from \cite{ER-tp}.}
\label{tptp-ap}
\end{figure}

The proton analyzing-power data of the Erlangen group \cite{ER-tp} cover the 
same angular range as the differential cross section; see fig. \ref{tptp-ap}. The $R$-matrix analysis
reproduces the data nicely, being slightly above the data for forward angles,
just missing the maximum value, and barely reaching the data in the
backward hemisphere. For the $NN$ interaction alone, the RRGM calculation is
always below the $R$-matrix analysis, closer to the data up to $70^\circ$, 
falls well below the maximum,
and also well below all backward data. This situation appears similar to the
notorious $A_y$ problem in the A = 3 systems  \cite{Ay} and also to missing 
the maximum polarization value in $p$ - $^3$He scattering \cite{BP}.
The full calculation, however, misses the forward data, but reaches the maximal value
of the $R$-matrix analysis at a somewhat smaller angle, and thus falls below
the backward polarization data. Note that we do not modify any $S$-matrix element,
contrary to the previous calculations \cite{HE4}.

\begin{figure}[h]
\centering
\includegraphics[width=8cm]{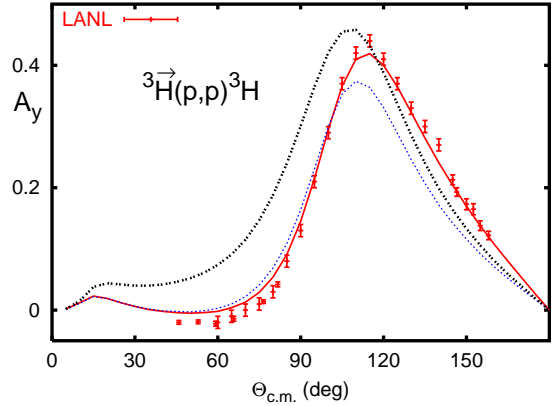}
\caption{(Color online) Triton analyzing power of the elastic scattering $^3$H($p,p$){$^3$H}
calculated at 3.2 MeV E$_{cm}$. The data at 3.21 MeV are from the Los Alamos 
group \cite{LANL-tp}.}
\label{tptp-at}
\end{figure}

For elastic proton-triton scattering, triton analyzing-power data also exist 
from the Los Alamos group \cite{LANL-tp} at a close-by energy in the same angular
range as the other data. The $R$-matrix just misses
the negative values at forward angles;
up to $80^\circ$ it is somewhat 
above the data, but then agrees nicely with them,
as shown in fig. \ref{tptp-at}. Using the AV18 alone
yields triton analyzing powers just  on top of the $R$-matrix results till its 
maximum, which is reached at a smaller angle and smaller value
than $R$-matrix and data. Afterwards
all calculated values are well below the data. Adding the UIX TNF yields much
too high polarization values at forward angles, reaches the maximal data at a
bit smaller angle,
and falls below the data and $R$-matrix fit at backward angles.
The large difference in 
the maximal proton and triton polarizations is caused by the rather large
$^3P_1$ to $^1P_1$ transition matrix element.

Summarizing the results for elastic proton-triton scattering, we see large
effects by adding UIX TNF, sometimes favorable, as for the maximal polarization
values, and sometimes adverse, as for the differential cross section and the forward
analyzing powers. Allowing in the calculation for $F$-waves and the $3^+$
$D$-wave always improved the agreement between $R$-matrix and RRGM results, but in
general the modifications were too small to display them clearly in figures.
Contrary to the previous calculation \cite{HE4}, we cannot identify a single
matrix element that causes the differences between the $R$-matrix analysis
and the RRGM calculation. Usually the moduli agree within a few percent and
the phase shifts within a few degrees. The only exception is the $^3P_0$
matrix element, which is small due to the strong coupling as discussed above.
Adding the TNF increases it by a factor three to 0.22 and reduces its phase
by $10^\circ$.
Furthermore, the results for this partial wave vary strongly close to the
energy considered; see fig. \ref{0m-au}. The general structure of the
differential cross-section and the two analyzing powers is already given by
the three triplet $P$ phase shifts and the $^3S_1$ ones. All the many others 
yield changes in the cross-section that are smaller than the difference between the $R$-matrix results and those of the full calculation. The differences in
the analyzing powers at forward angles and around the maximum
comes mainly from the slightly differing $^3P_2$ matrix element and, to a lesser extent,
from the size of the $^3P_0$ $S$-matrix element (the other two are so close
that no differences are visible). Since the the maximal polarization values
are reached by the full calculation also at higher energies, we do not see
indications here of an "$A_y$ problem", but we note that the
forward analyzing powers are missed.

\begin{figure}[h]
\centering
\includegraphics[width=8cm]{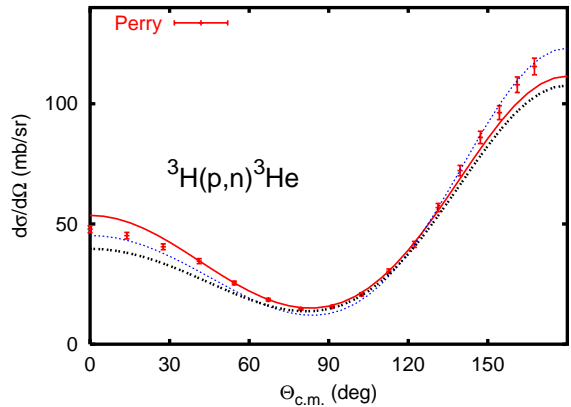}
\caption{(Color online) Differential cross section for the reaction $^3$H($p,n)^3$He
calculated at 3.0 MeV E$_{cm}$. The data at 3.08 MeV are from
Perry et al.\cite{Perry}.}
\label{tph-x}
\end{figure}

The differential cross section for the reaction  $^3$H($p,n)^3$He is shown in
Fig. \ref{tph-x}. The $R$-matrix analysis is somewhat above the very forward data
and a bit below them at backward angles. The calculation for the AV18 potential
alone reproduces the data very nicely, being only slightly below at forward angles.
Adding the TNF destroys again the agreement, by loosing strength at forward
and backward angles. All the large matrix elements agree between analysis and
the calculations, with only the exception of the $^3P_1$ matrix element, which
is above 0.2 in the $R$-matrix and in the full calculation, but only 0.04 for
the AV18 alone. The modulus of this matrix element rises rapidly with energy from
threshold to about 2.5 MeV $E_{cm}$ above the triton-proton threshold in
analysis and the full calculation,
falling to a minimum near 5 MeV, and a gentle increase afterwards. The AV18
calculation, however, yields the maximum at 2 MeV, the minimum very
close to zero just above 3 MeV, and a rather rapid increase afterwards. These
rapid variations of this matrix element are the main reason for the rather strong
energy dependence of the observables of the $^3$H($p,n)^3$He reaction. In 
addition, the moduli of the $D$-wave matrix elements increase rapidly with
energy, thus leading to major changes within 200 keV, especially in 
polarization observables not displayed here.
The final results are thus determined by many small matrix elements that
change rapidly with energy,
rather than by the large $0^-$ and $0^+$ ones at the unitary limit. Unfortunately,
all the polarization data are concentrated below 3.5 MeV, and none exist around
7.5 MeV, where all $D$-waves belong to the large matrix elements in the analysis.

The time-reversed reaction $^3$He(n,p)$^3$H is used at low energies as a neutron
standard reaction. These integrated cross sections are reasonably well
reproduced by the current calculation (see \cite{SCAT-LE}, where an early
version of the present calculations is reported). The main emphasis of 
\cite{SCAT-LE}, however, was the determination of the spin-dependent scattering
lengths $a_s$
of $^3$He-neutron elastic scattering. In the meantime a new measurement
of the coherent neutron scattering length exists \cite{Snow} in addition to the
recent measurement of the incoherent one \cite{Zimmer}. Due to numerical problems
we could hardly go down to 1 keV in the early
RRGM calculations in order to determine
the complex scattering lengths. Special measures had to be taken to extract them
from the calculated $S$-matrix elements. Increasing the numerical stability by the
procedure described in the beginning, we can now go down safely to 0.1 keV and
check the extrapolation at even lower energies. The real part of $a_s$ can now
be calculated via the standard expression $a = \tan \delta /k$, whereas the
imaginary part of $a_0$ still needs the expression given in \cite{SCAT-LE}.
The calculated results for $a_1$ are within the errors given in \cite{SCAT-LE}.
Since the measured coherent and incoherent scattering lengths are linear combinations
of $a_0$ and $a_1$, and therefore always need the input of other data, we compare
in tab. \ref{scat} the calculated values with the extracted data.

\begin{table}
\centering
 \caption{\label{scat} Comparison of experimental and
 calculated real and imaginary scattering lengths (in fm) for
 the potential models used}
 \vskip 0.2cm
 \begin{tabular}{c|c|c|c|c}
 \hline\noalign{\smallskip}
 potential & \multicolumn{2}{c|}{$a_{0}$} & \multicolumn{2}{c}{$a_{1}$} \\
     & $\Re$ & $\Im $ & $\Re$  &  $ \Im  $ \\
     \noalign{\smallskip}\hline\noalign{\smallskip}
     AV18    & 7.776(1)& -5.019(1) & 3.447(1) & -0.0066(1) \\
     AV18 + UIX & 7.622(1) & -4.095(1) & 3.311(1) & -0.0051(1) \\
     $R$-matrix & 7.400(3) & -4.449(1) & 3.286(6)& -0.0012(2)\\
     exp.    & 7.370(58) \cite{Zimmer}& -4.448(5) \cite{Kohl} & 3.278(53)
\cite{Zimmer}& -0.001(2) \cite{Kohl}\\
     exp.    & 7.456(20) \cite{Snow}&  & 3.363(13)
\cite{Snow}& \\
     \noalign{\smallskip}\hline
     \end{tabular}
 \end{table}

\begin{figure}[h]
\centering
\includegraphics[width=8cm]{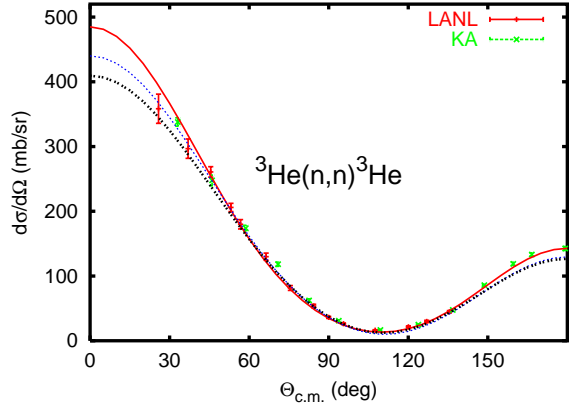}
\caption{(Color online) Differential cross section for the elastic scattering $^3$He(n,n)$^3$He
calculated at 6.0 MeV E$_{cm}$. The data are from Drosg \cite{Drosg-he3n} at
5.93 MeV and from the Karlsuhe-group \cite{Klages-he3n} at 6.0 MeV.}
\label{he-x}
\end{figure}

We note in passing that the calculated coherent scattering length for the full
calculation agrees perfectly with the new measurement \cite{Snow}. Since
the calculation, however, yields the spin-zero and spin-one parts separately,
this agreement has to be considered fortuitious. The most recent measurement
puts the spin-dependent scattering lengths well outside the older error bars.

\begin{figure}[h]
\centering
\includegraphics[width=8cm]{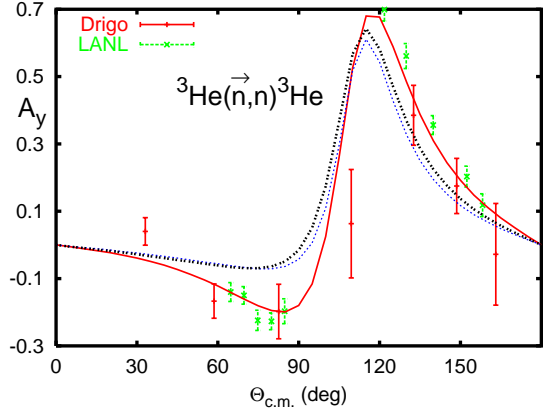}
\caption{(Color online) Neutron analyzing power for the elastic scattering $^3$He(n,n)$^3$He
calculated at 6.0 MeV E$_{cm}$. The data are from Drigo \cite{Drigo} and
LANL \cite{Lisowski} at 5.85 MeV and 6.0 MeV respectively.}
\label{he-ay}
\end{figure}

For elastic $^3$He-neutron scattering, cross section and analyzing-power data
do not exist at the energy used for triton-proton  scattering. In figs. 
\ref{he-x} and \ref{he-ay} respectively, we display such data, measured close to 6 MeV
E$_{cm}$. The $R$-matrix analysis reproduces the cross section and
analyzing-power data very well.
The RRGM calculations do not reach up to the Karlsruhe cross sections
\cite{Klages-he3n} with their tiny errors at backward angles. Also the negative
analyzing powers below $90^\circ$ are not reached, the maximum is slightly missed,
and they fall below the data at backward angles.
Despite the similarity of the results
with and without TNF, individual matrix elements turn out to be quite different in
phase and/or modulus, but some changes reduce the polarization, others
increase it, resulting in an almost zero net change.
The differences between $R$-matrix analysis and the calculations are mainly due to
the larger triplet-P phase shifts, shown in figs. \ref{2m-auF}, \ref{1m-au}, and \ref{0m-au},
which increase the cross section at forward and backward angles and yield
strongly negative analyzing powers near $90^\circ$.
Due to the higher energy, the $^3P_0$
matrix element has increased by a factor of 2 compared to triton-proton scattering,
and therefore lost some of its sensitivity. The positive $^3D_2$ phase shift
from the $R$-matrix, fig. \ref{2p-au},
compensates for this large $P$-wave by reducing the maximal analyzing power and the
$180^\circ$ cross section. As in the triton-proton scattering, the three
triplet $P$ phase shifts and the $^3S_1$ one determine the overall structure
of cross section and analyzing power, but here, due to the higher energy, the
effects of the other matrix elements are larger for the cross section. For the
analyzing powers the effectsof these other matrix elements are visible only in
the fall-off from the maximum. The maximal
polarization values are almost reached by the calculations.

\begin{figure}[h]
\hspace{-0.5cm}
\includegraphics[width=8cm]{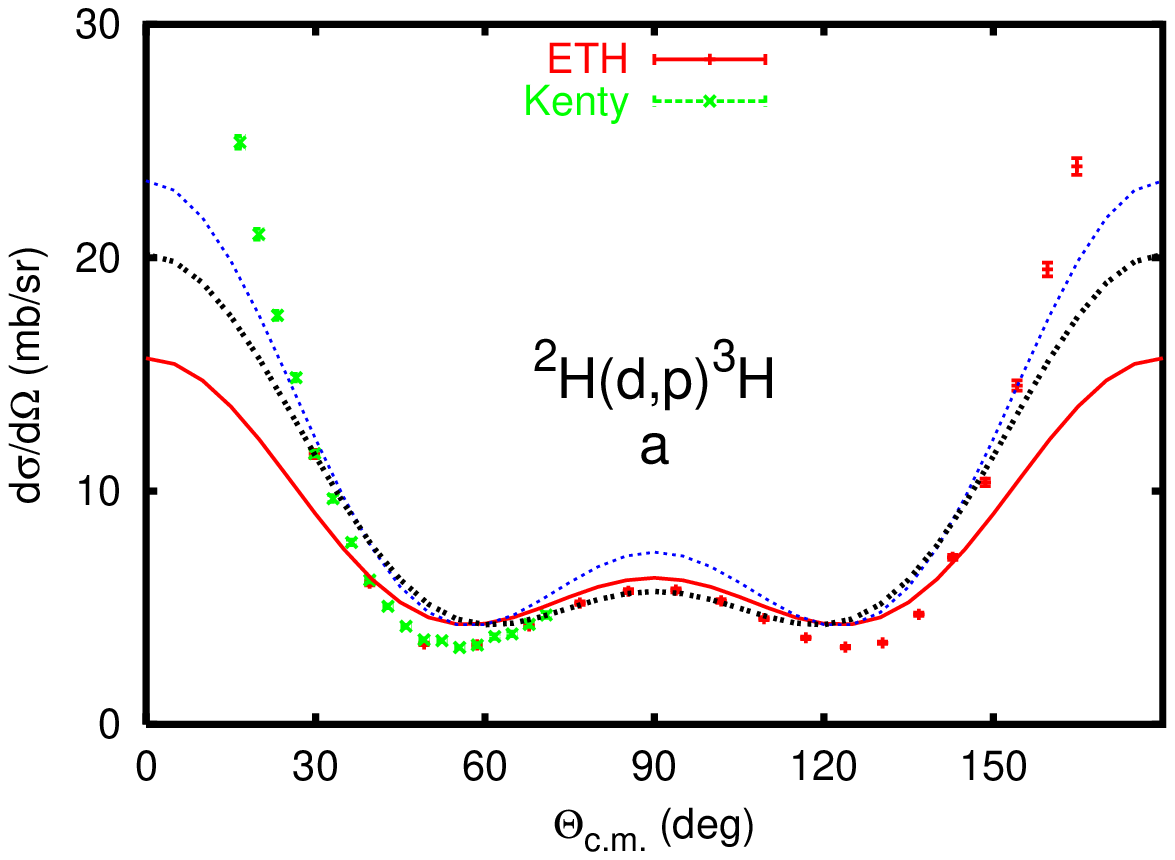}
\hspace{0.5cm}
\includegraphics[width=8cm]{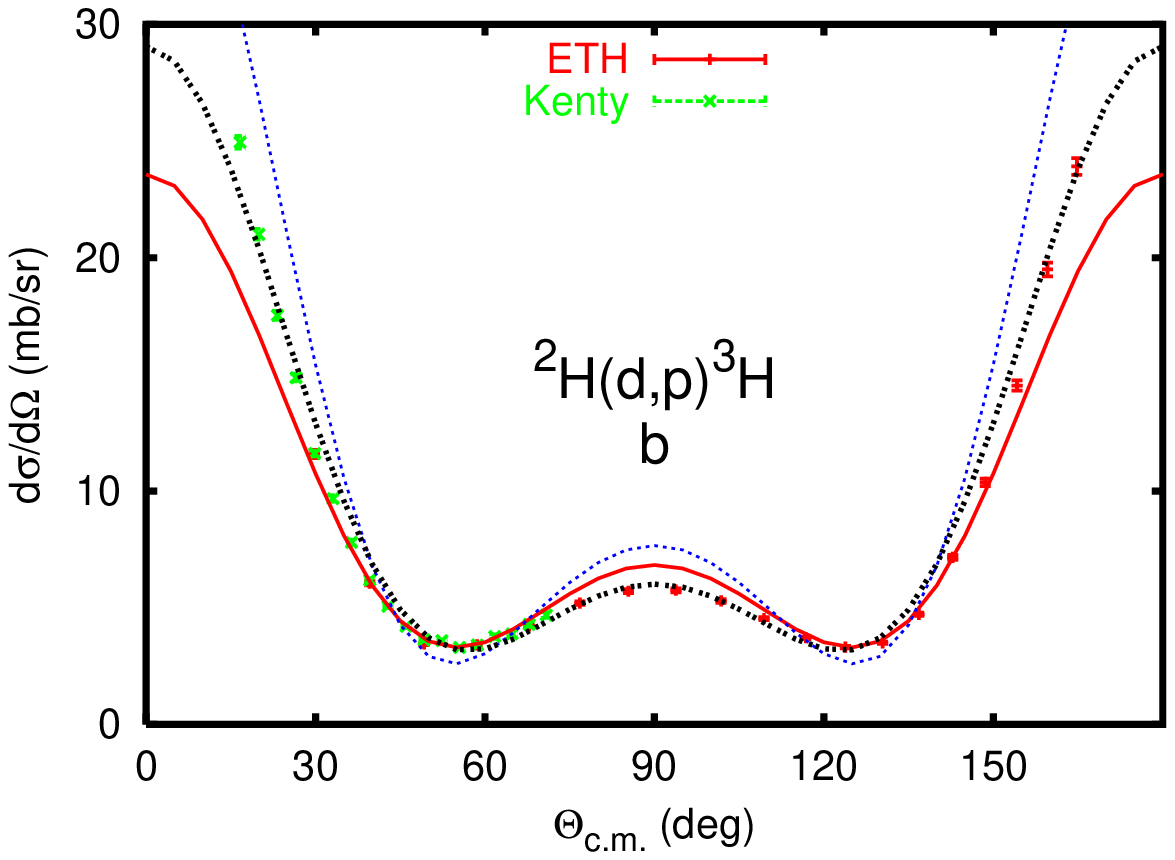}
\caption{(Color online) Differential cross section for the reaction $^2$H($d,p$)$^3$H
calculated at 2.0 MeV E$_{cm}$. 
The data are from the Z\"urich group \cite{Grueb} and from the Kentucky group
\cite{Kenty} at 2.0 MeV.
For AV18 alone the energy is chosen in such a way that  it agrees in the exit 
channel with the experimental one.
$F$-waves are not taken into account in a, but in b.
}
\label{ddp-x}
\end{figure}

Let us now discuss the deuteron induced reactions. In fig. \ref{ddp-x} the
$^2$H($d,p$)$^3$H differential cross section is displayed without and with $F$-waves
taken into account. Without $F$-waves only the pure $NN$-only calculation comes
close to the data, whereas with $F$-waves the data are rather nicely 
reproduced. The $R$-matrix analysis underestimates the cross section at the 
extreme forward and backward angles and overestimates it around 90 degrees.
For the AV18 potential alone, the cross section is mostly overpredicted, but
including the TNF yields results on top of the data. 
Due to the identical particles in the entrance channel the cross section has to
be symmetric about $90^\circ$, which is not quite true for the two different
data sets at the extreme angles. The forward Kentucky data are about 10 percent
above the corresponding Z\"urich data at backward angles, which is revealed by
comparing to the full calculation.
We postpone the discussion of the effects of individual transition matrix elements
until we have compared also all analyzing powers for this reaction, in order to
avoid unneccessary repetition.
 
\begin{figure}[h]
\hspace{-0.5cm}
\includegraphics[width=8cm]{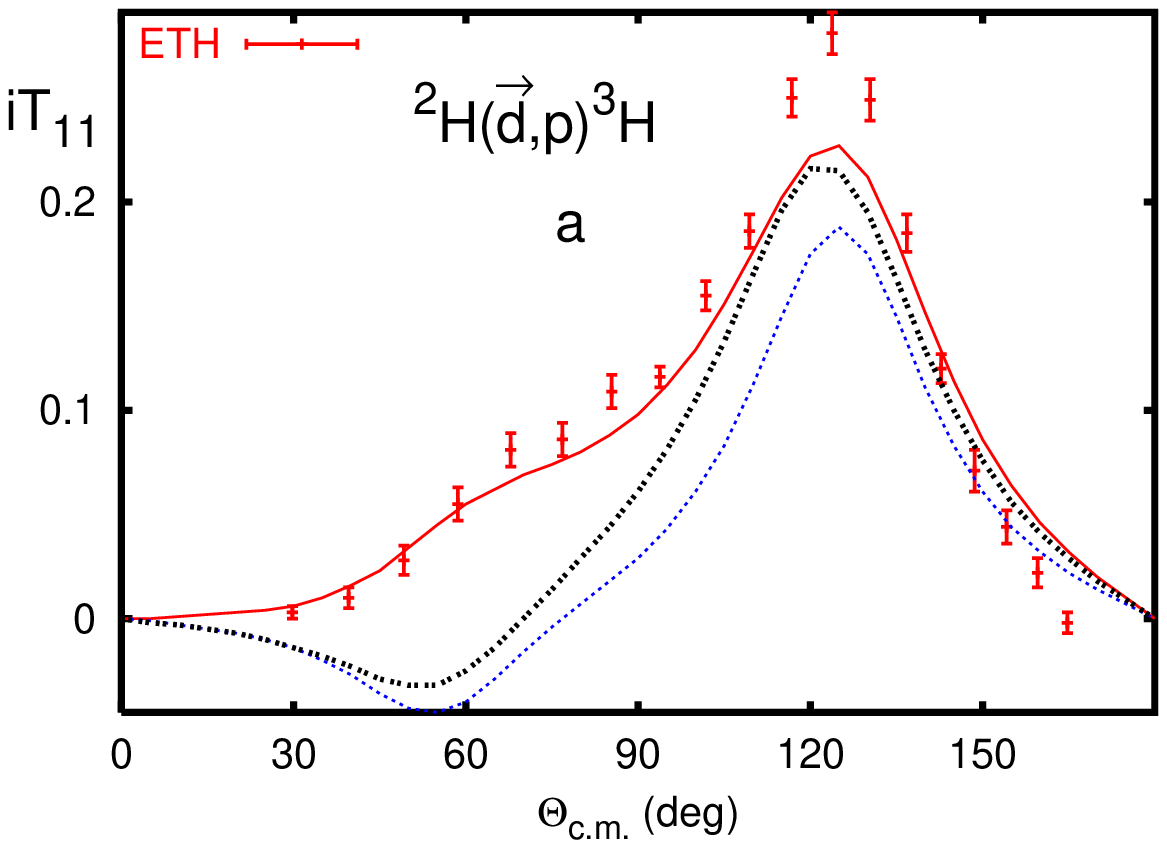}
\hspace{0.5cm}
\includegraphics[width=8cm]{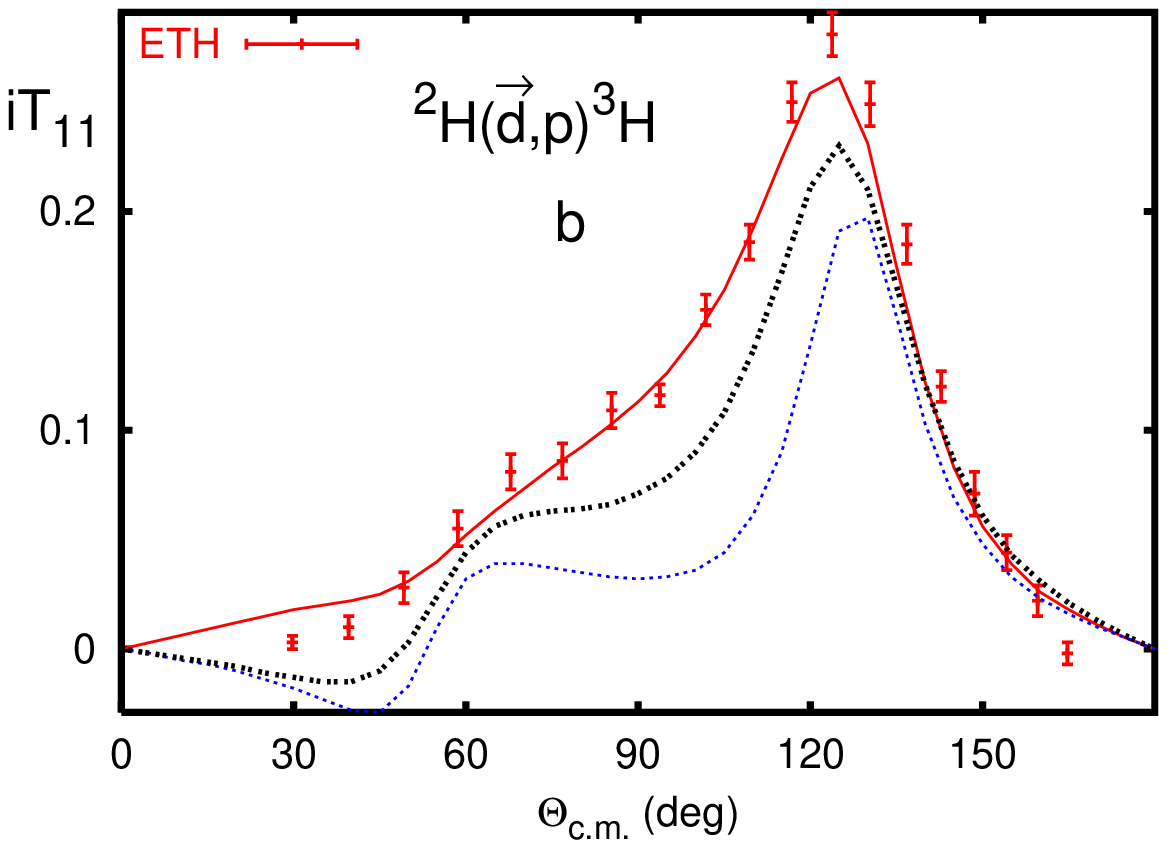}
\caption{(Color online) Same as fig. \ref{ddp-x} but for the analyzing power $iT_{11}$ of the 
reaction $^2$H($d,p$)$^3$H
calculated at 2.0 MeV E$_{cm}$. The data at 2.0 MeV are from the
Z\"urich group \cite{Grueb}, as all the following polarization data.}
\label{ddp-it11}
\end{figure}

The vector analyzing power is displayed in fig. \ref{ddp-it11}. The $R$-matrix
analysis does a nice job of reproducing most of the data. Without $F$-waves, the
forward hemisphere data are well reproduced, the maximum is not reached,
and the very small values above 150$^\circ$ are missed. Including $F$-waves, the
data are very well reproduced, except for the zero values at forward and backward
angles, and maybe also in the maximum. The RRGM calculations yield
negative polarizations for forward angles, contrary to data; adding TNF makes the
results always more positive, as do the additional $F$-waves in the forward
hemisphere. So the full calculation
comes closest to the data, but cannot be called a good description of them.
The AV18 calculation without $F$-waves yields only qualitative agreement.

\begin{figure}[h]
\hspace{-0.5cm}
\includegraphics[width=8cm]{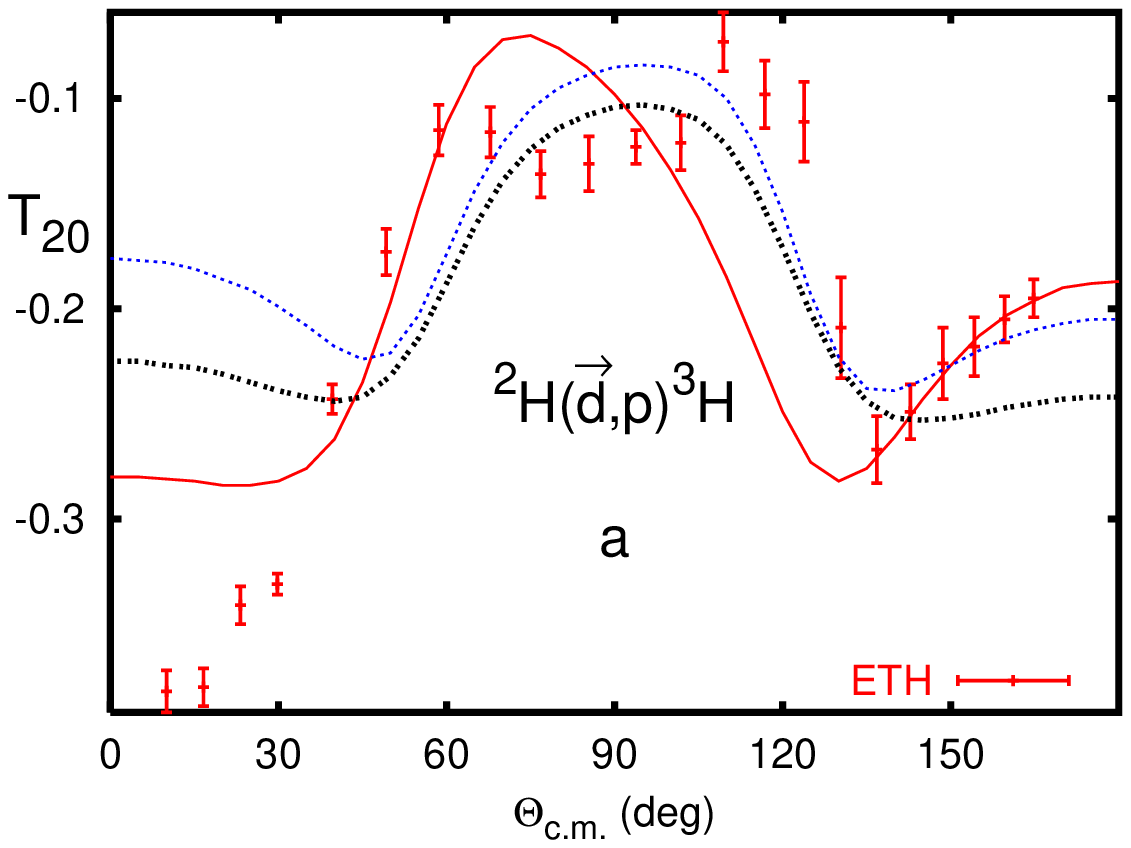}
\hspace{0.5cm}
\includegraphics[width=8cm]{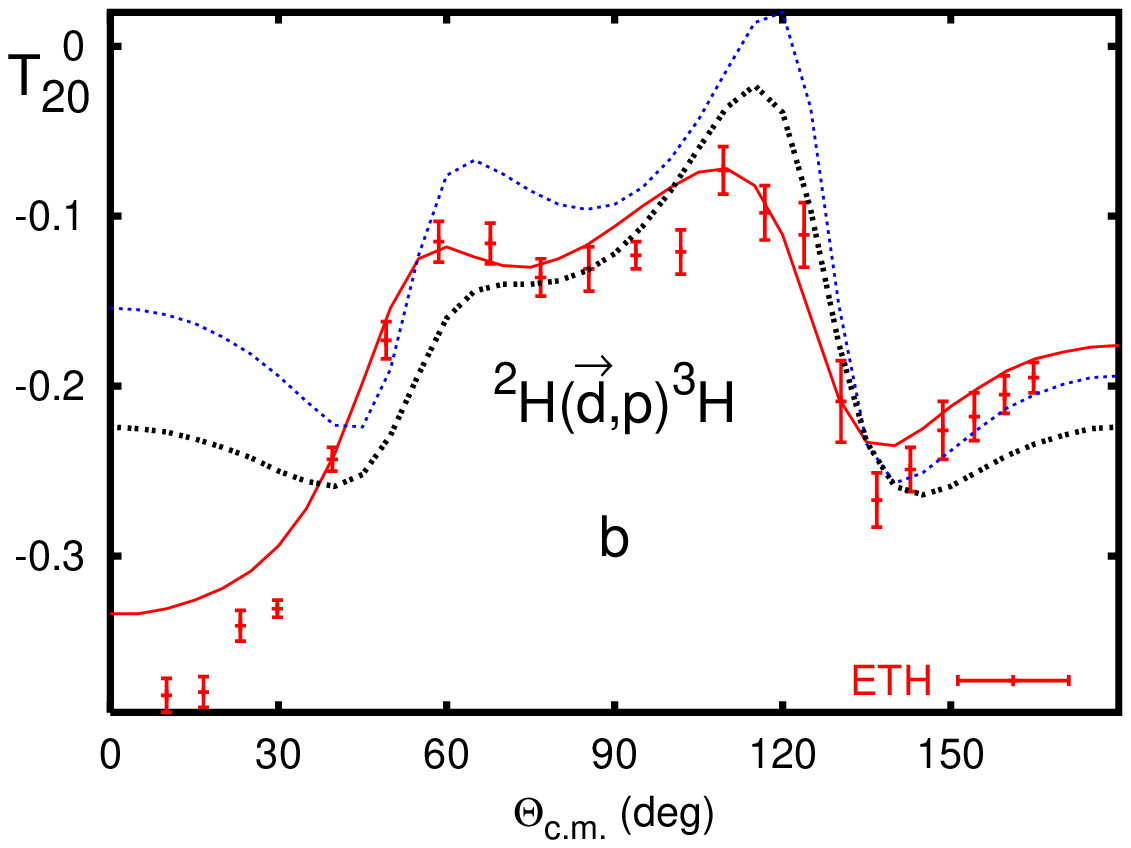}
\caption{(Color online) Same as fig. \ref{ddp-it11}, but for the analyzing power $T_{20}$ of 
the reaction $^2$H($d,p$)$^3$H}
\label{ddp-t20}
\end{figure}

The $T_{20}$ analyzing power is displayed in fig. \ref{ddp-t20}. Without $F$-waves
the double hump structure of the data is completely missed, and also the very
negative values below 30 degrees. Including $F$-waves, the $R$-matrix analysis
reproduces the data very well except for the forward angles. Both RRGM calculations
miss the forward data totally. The AV18 alone overestimates the double hump
structure, and including TNF improves the reproduction of the data somewhat.

\begin{figure}[h]
\hspace{-0.5cm}
\includegraphics[width=8cm]{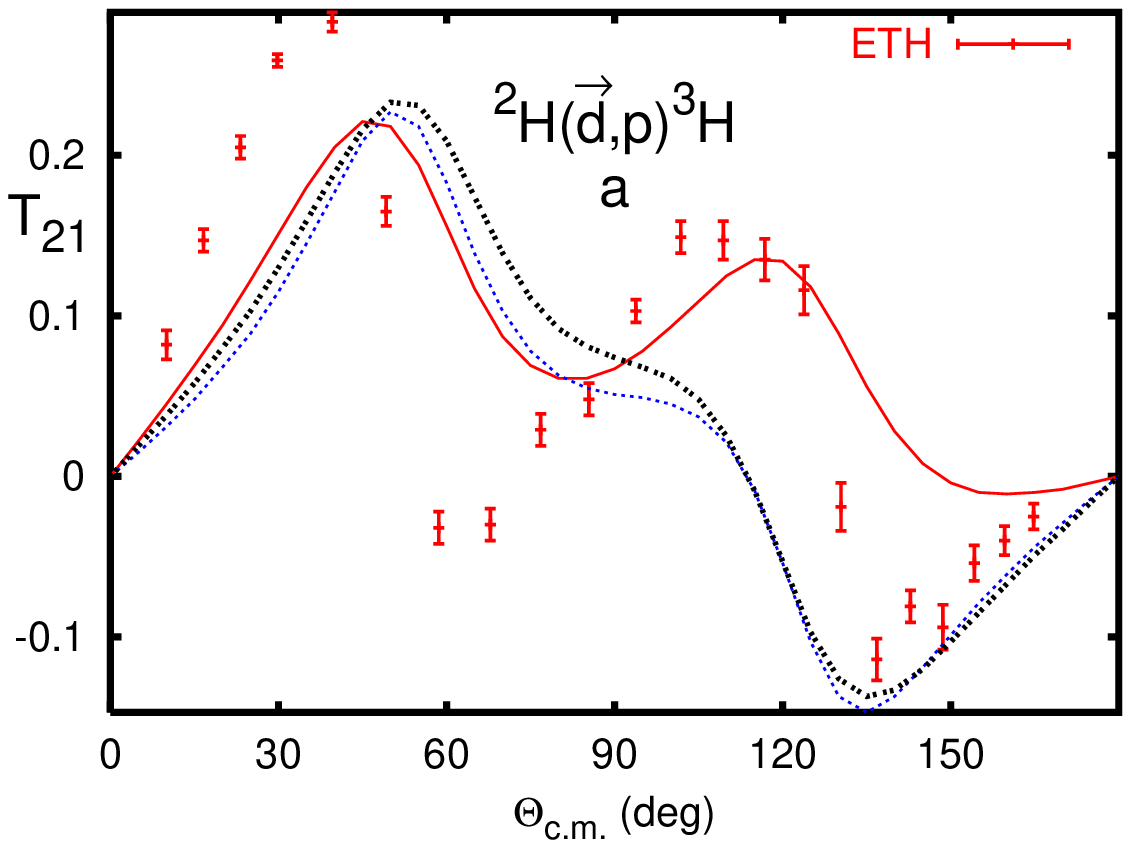}
\hspace{0.5cm}
\includegraphics[width=8cm]{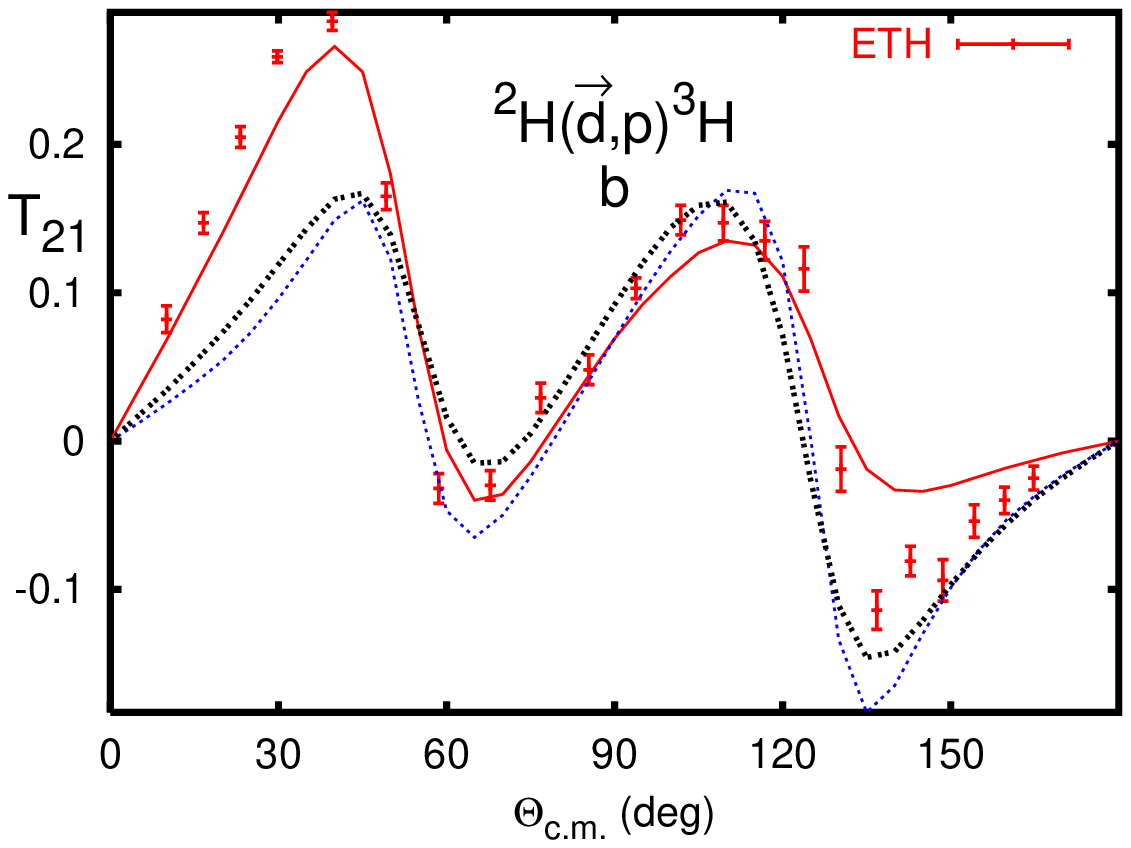}
\caption{(Color online) Same as fig. \ref{ddp-it11}, but for the analyzing power $T_{21}$ of 
the reaction $^2$H($d,p$)$^3$H}
\label{ddp-t21}
\end{figure}

In fig. \ref{ddp-t21} we display the tensor analyzing power $T_{21}$. Without
$F$-waves, only the $R$-matrix results are similar to the data. Below 90 degrees
the RRGM results are close to the $R$-matrix ones, being almost antisymmetric.
The RRGM calculations show almost no effect of the TNF.
The $R$-matrix analysis including $F$-waves reproduces the data well with the
exception of around $135^\circ$, where it becomes not negative enough. The RRGM
calculation reaches these data, but misses the steep rise from zero to the
first maximum. The overall agreement is satisfactory.

\begin{figure}[h]
\hspace{-0.5cm}
\includegraphics[width=8cm]{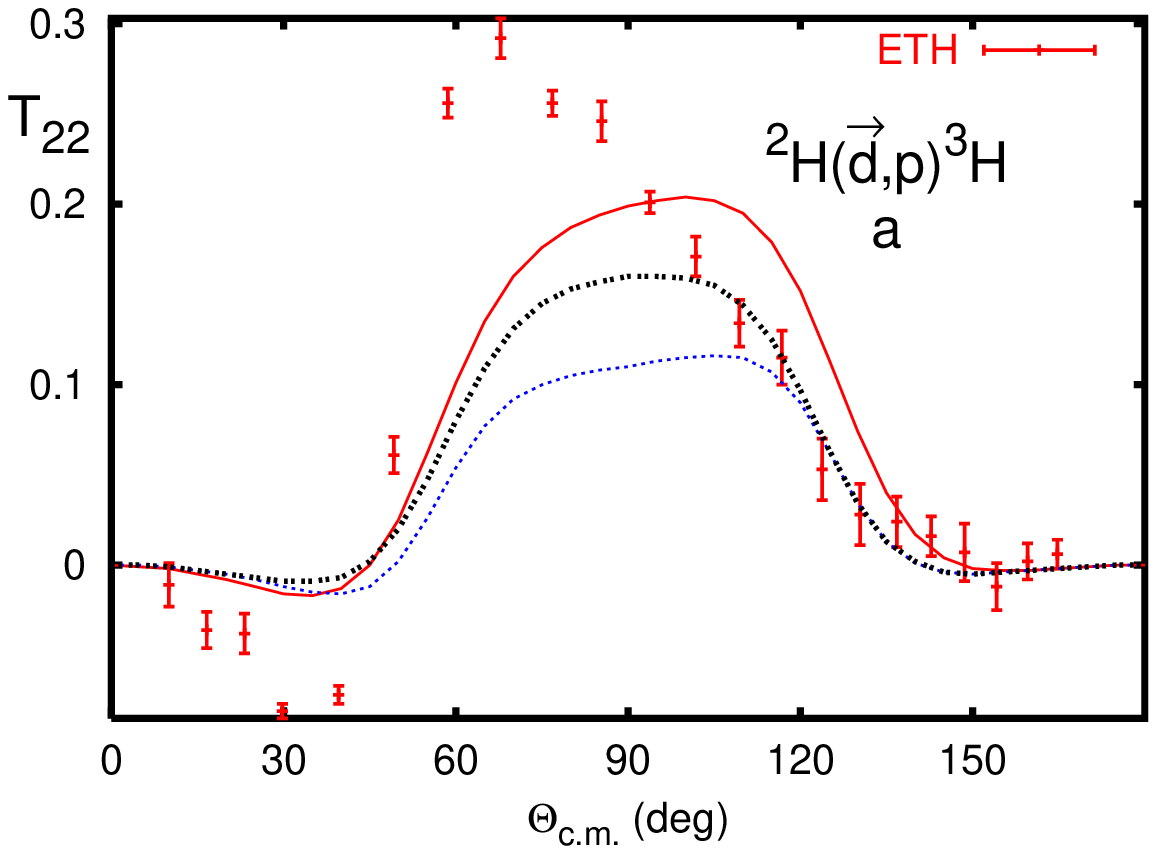}
\hspace{0.5cm}
\includegraphics[width=8cm]{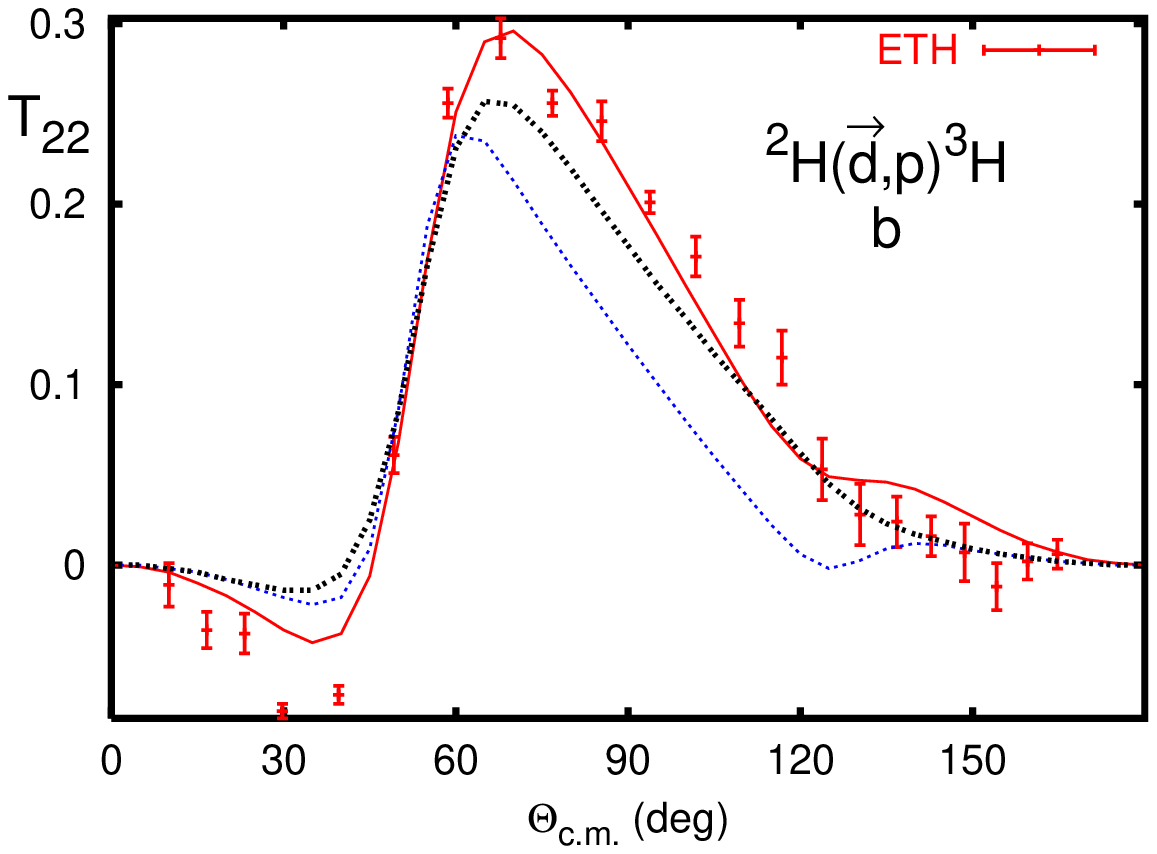}
\caption{(Color online) Same as fig. \ref{ddp-it11}, but for the analyzing power $T_{22}$ of 
the reaction $^2$H($d,p$)$^3$H}
\label{ddp-t22}
\end{figure}

Without $F$-waves, the tensor analyzing power $T_{22}$ shows again no qualitative
agreement with the data, as shown in fig. \ref{ddp-t22}.
The $R$-matrix  analysis and RRGM are all similar in shape;
this holds also true when including $F$-waves. Again the $R$-matrix does a nice job,
only underestimating the negative polarizations around 30 degrees. In this
angular range the RRGM results are even less negative. Both calculations miss
the height of the maximum, the $NN$ force also the angle, and hence, also the 
falloff to zero, whereas including TNF yields reasonable agreement.

Let us now compare the results of the $R$-matrix analysis and the RRGM calculations
in detail. All numerical work yields the $^3P_1$ and $^1D_2$ transition matrix 
elements as the largest ones,
with the AV18 calculation creating the largest values;
adding TNF, the typical loss is 10 percent, and the $R$-matrix ones are another
10 percent smaller. These ratios already take care of the forward-backward 
cross section behavior, as shown in fig. \ref{ddp-x}. The next larger matrix
elements are much smaller, typically less than half the largest one.
In the $R$-matrix analysis come the  $^5S_2 \rightarrow {^3D_2}$  and then the 
$^1S_0$ transitions, whereas the RRGM yields the $^1S_0$ matrix element as the next larger.
Adding the $^1S_0$  matrix elements keeps the cross sections as expected and
still yields zero vector analyzing power since there is no second channel with
which to 
interfere.  Adding the   $^5S_2 \rightarrow {^3D_2}$ transition sets
the scale for the maximum of the vector analyzing power. The $R$-matrix 
analysis gives the
highest value, then the full RRGM calculation, and then $NN$ forces only, 
as seen in fig. \ref{ddp-it11}.
Also this transition allows the tensor analyzing power $T_{20}$ to approach
its ultimate value around zero degrees,
with the $R$-matrix giving the most negative value, which becomes 
somewhat more negative with increasing angle, before it reduces to the first maximum.
The calculation including TNF yields a less negative value at $0^\circ$,
becomes much more negative with increasing angle, until it also reduces to
smaller values.
The two RRGM calculations are quite similar to each other, with the AV18
calculation yielding always less negative values.
Adding the next largest matrix element, which is the  $^5D_1 \rightarrow {^3D_1}$
one for the $R$-matrix, leads to the saw-tooth structure for $T_{21}$
(see fig. \ref{ddp-t21}a), and the slightly negative values of $T_{22}$ at small
angles, followed by an increase, similar to fig. \ref{ddp-t22}a.
In the RRGM, this matrix element is only a quarter of the $R$-matrix size;
therefore it plays no essential role. At this stage, however,
the $^3P_2$ has to be taken
into account. This matrix element leads to negative $iT_{11}$ values similar to
Fig. \ref{ddp-it11}a, a behavior that is not changed by additional matrix
elements. 
The resulting tensor analyzing powers $T_{21}$ and $T_{22}$ are
still far from any qualitative structure of the data. Also adding the
$^5S_2 \rightarrow {^1D_2}$ transition
does not change the situation. Only when adding the
$^3F_4$ and $^3F_3$ matrix elements, the RRGM reproduces qualitatively the data.
The rest of the agreement of the final figs. \ref{ddp-it11}, \ref{ddp-t20}, \ref{ddp-t21},
and \ref{ddp-t22} is due to the interplay of many more small matrix elements.
One additional feature of the small matrix elements deserves mentioning: The
$^5D_3 \rightarrow {^3D_3}$ matrix element in the RRGM calculation has the same
magnitude, but opposite phase of the corresponding matrix elements coupled to
total J of one or two. This sign change is in accordance with a dominating
effective tensor force for this transition; however, the agreement with the
polarization data becomes worse. In the $R$-matrix analysis this matrix element
is essentially zero, and thus has no effect.

\begin{figure}[h]
\centering
\includegraphics[width=8cm]{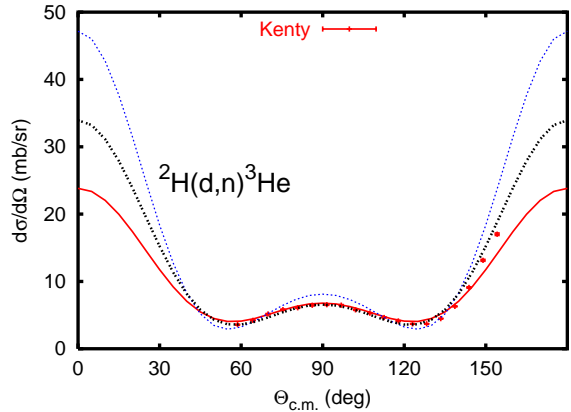}
\caption{(Color online) Differential cross section for the reaction $^2$H($d,n$)$^3$He
calculated at 2.0 MeV E$_{cm}$ with all matrix elements taken into account.
For AV18 alone the energy is chosen that in the exit 
channel it agrees with the experimental one. The data from the Kentucky group
\cite{Kenty} are at 2.0 MeV.}
\label{ddn-x}
\end{figure}

For the charge conjugate reaction $^2$H($d,n$){$^3$He}, data exist from the Kentucky
group  \cite{Kenty} and the OSU group \cite{Dries} 
at the same energy. Since charge symmetry is rather good,
all data and calculations for the reaction  $^2$H($d,n$){$^3$He} should be
similar to those of  $^2$H($d,p$){$^3$H}, except for the small change in the
energy of the exit channel. Therefore, we present only the results for the 
calculations including all channels.
The OSU group measured the cartesian components of the analyzing powers;
hence, for $A_{xx}$ there is no direct counterpart in the proton channel, and
we present also the results without $F$-waves taken into account. In fig. 
\ref{ddn-x} we compare the differential cross section data to the various
calculations. As in the proton channel, the $R$-matrix analysis underestimates
the backward data, the calculation using AV18 alone overestimates them, and including
TNF brings the results in agreement with the data. Unfortunately the angular
range of the data is rather limited.


\begin{figure}[h]
\centering
\includegraphics[width=8cm]{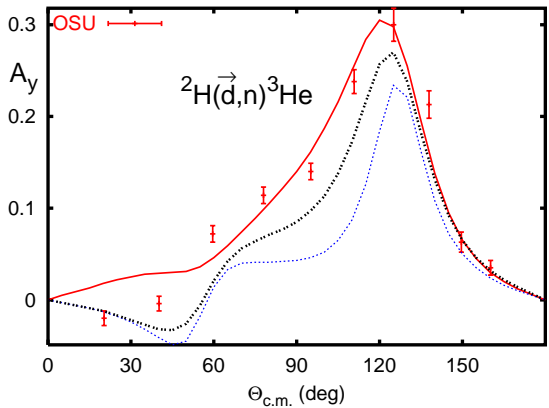}
\caption{(Color online) Same as fig. \ref{ddn-x}, but for the vector analyzing power $A_y$.
The data are from the Ohio State group \cite{Dries} at 2.0 MeV, as
are all the following polarization data.}
\label{ddn-ay}
\end{figure}

In fig. \ref{ddn-ay} we display the vector analyzing power $A_y$. The results
of all calculations are quite similar to the proton calculations shown in 
Fig. \ref{ddp-it11}b. The data of 
the OSU group \cite{Dries} are slightly negative at forward angles, which the
$R$-matrix analysis does not follow.


\begin{figure}[h]
\centering
\includegraphics[width=8cm]{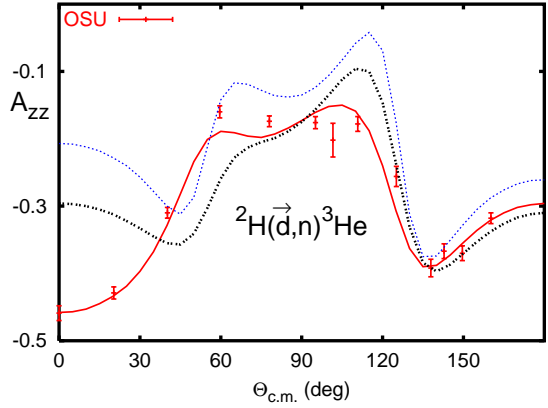}
\caption{(Color online) Same as fig. \ref{ddn-ay}, but for the tensor analyzing power $A_{zz}$.}
\label{ddn-azz}
\end{figure}

The $R$-matrix analysis reproduces the tensor analyzing power $A_{zz}$ in fig.
\ref{ddn-azz} very well, also at the extreme forward and backward angles. As
in the proton channel, the RRGM calculations do not reach the minimum at zero 
degrees. Also the structure before the second maximum is not well reproduced.
Some of the disagreement is due to the $^5D_3 \rightarrow {^3D_3}$ matrix element,
as discussed above.


\begin{figure}[h]
\hspace{-0.5cm}
\includegraphics[width=8cm]{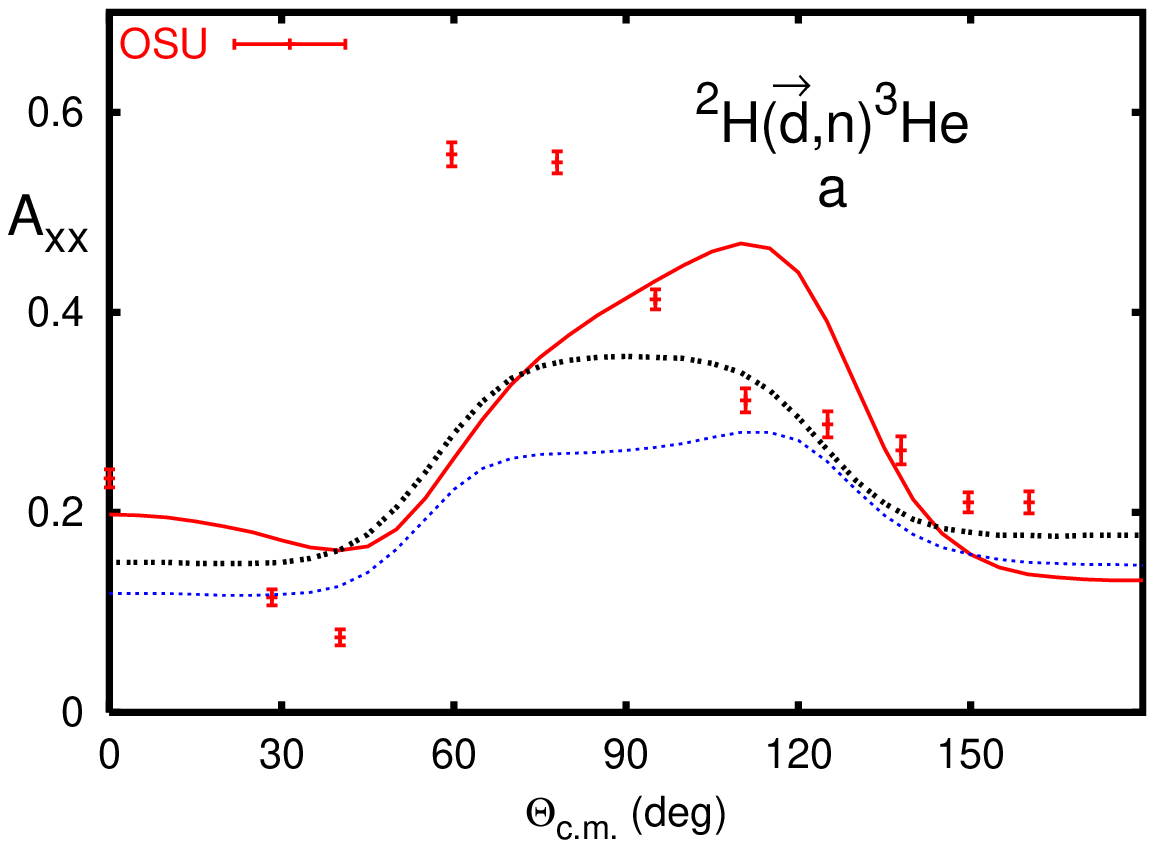}
\hspace{0.5cm}
\includegraphics[width=8cm]{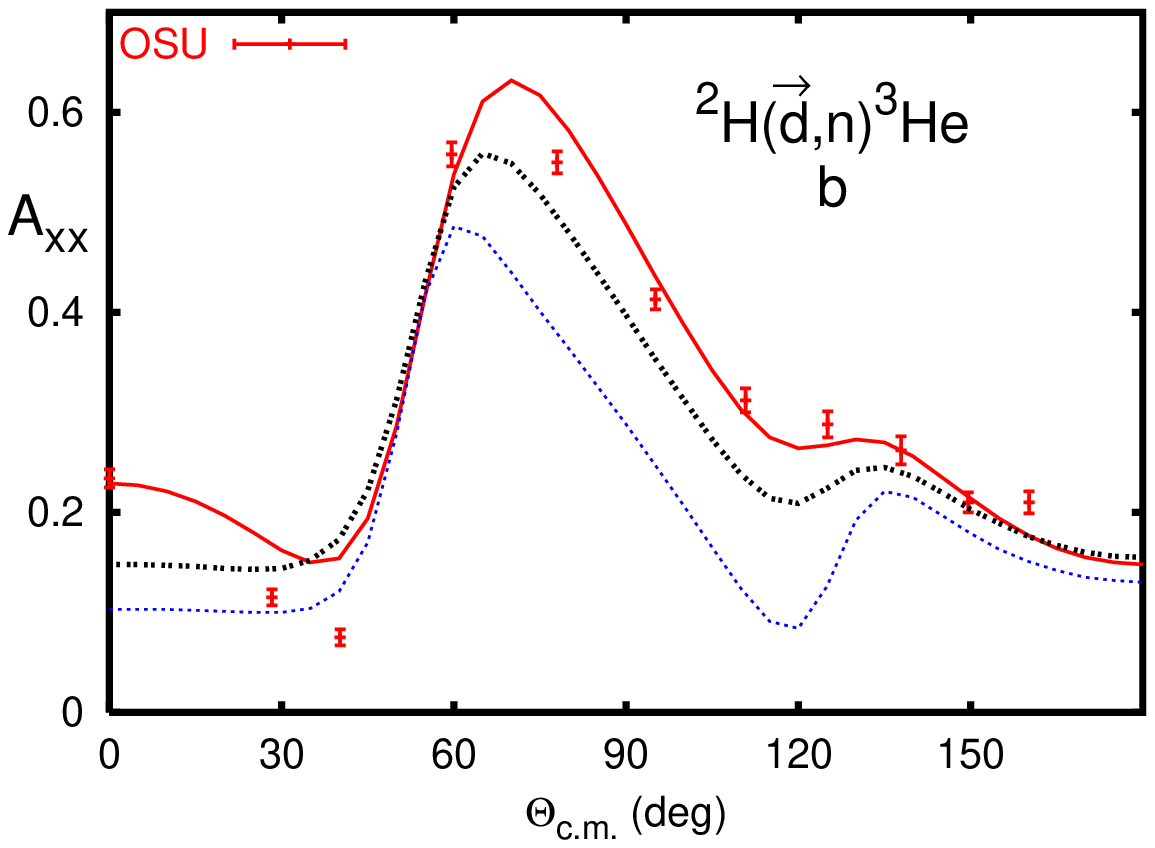}
\caption{(Color online) Same as fig. \ref{ddn-ay}, but for the tensor analyzing power $A_{xx}$,
$F$-waves are taken into account in (b), but not in (a).}
\label{ddn-axx}
\end{figure}

Since the tensor analyzing power $A_{xx}$ has no direct counterpart in the
proton channel, we present in fig. \ref{ddn-axx} the results with and without
$F$-waves. Without $F$-waves, the structure of the calculation does not reproduce the 
data, and only the $R$-matrix analysis gives qualitative agreement up to $60^\circ$.
Including $F$-waves the $R$-matrix reproduces the data well, except close to 
thirty degrees. The RRGM calculations agree qualitatively with the data, with
major deficiencies below $30^\circ$ and around $120^\circ$. The full calculation
agrees better overall.


\begin{figure}[h]
\centering
\includegraphics[width=8cm]{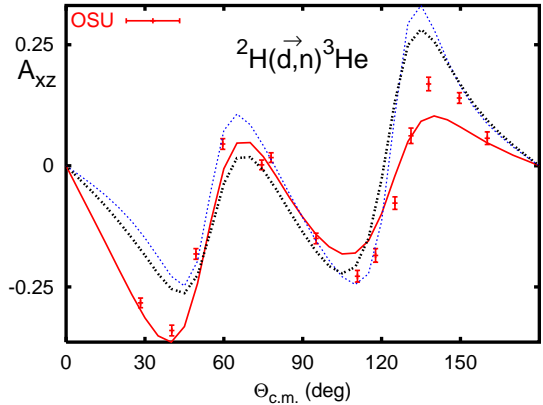}
\caption{(Color online) Same as fig. \ref{ddn-ay}, but for the tensor analyzing power $A_{xz}$.        }
\label{ddn-axz}
\end{figure}


As for $T_{21}$ of the proton channel, the saw-tooth structure of the tensor analyzing
power $A_{xz}$ is well reproduced by all the calculations. The $R$-matrix has some
difficulties around $150^\circ$, and 
the RRGM results differ there and also around $45^\circ$. Note the sign difference
between figures \ref{ddn-axz} and \ref{ddp-t21} due to the definition of $A_{xz}$
and $T_{21}$.

Considering all results for the $^2$H($d,n$){$^3$He} reaction, we find very close
similarity with the charge conjugate proton channel. Some observables are
better reproduced by the $R$-matrix analysis in the proton channel, 
others in the
neutron channel. The parameter-free RRGM calculations do not reach a similar
agreement for the analyzing powers. The transition matrix elements follow the
same pattern as for the proton channel, although the small matrix elements
usually have a somewhat larger effect.

One of the significant differences noted earlier between the $R$-matrix fit 
and the measurements was for the $d$ - $d$ reaction differential cross sections
at forward and backward angles, even with the $F$-wave transitions included.  
In comparing with the RRGM calculations, which predict those cross sections 
much better, it appears that this difference comes primarily from the 
interference of the $^1S_0$ and $^1D_2$ transition matrix elements.   
The $^1S_0$ transition obtained from the $R$-matrix fitting has an interference
effect between levels at low energies that does not show up in the RRGM 
calculations.  Because of this interference effect, the phase of the transition
element changes sign and gives the opposite interference behavior with the
$^1D_2$ matrix element at most energies, compared with the RRGM calculations.
This results in an important component of the ``$P_2$" behavior of the 
differential cross section having the opposite sign with respect to the 
calculations, giving cross sections that are too high at 90 degrees, and too 
low at forward and backward angles. 

Earlier in the analysis, it was thought that the deficiency was due to the lack
of higher partial waves.  However, the comparison to the RRGM calculations 
reveals that the problem is likely this anomalous interference pattern in 
the $0^+$ levels of the fit.  We speculate that this behavior arose in order 
to give a rapid rise in the $S$-wave reaction cross sections at low energies,
and is symptomatic of the $d$ - $d$ channel radius (7 fm) having too small a value.

\begin{figure}[h]
\centering
\includegraphics[width=8cm]{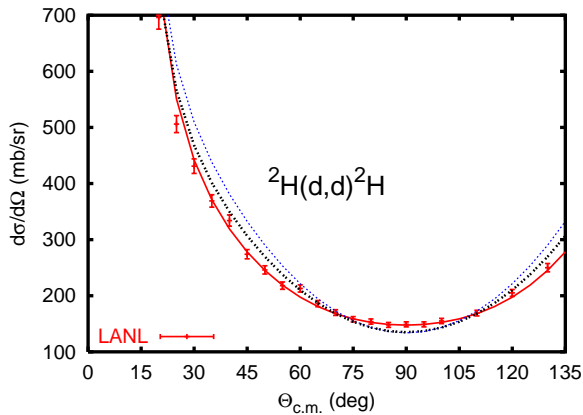}
\caption{(Color online) Differential cross section for elastic deuteron-deuteron scattering
calculated at 3.0 MeV E$_{cm}$ including $F$-waves. The data are from \cite{Brolley} at 3.0 MeV.}
\label{dddd-x}
\end{figure}

Let us now discuss the last two-body process, the elastic deuteron-deuteron
scattering. Due to the identical bosons in entrance and exit channel, the
vector analyzing power $iT_{11}$ and the tensor analyzing power $T_{21}$ are
antisymmetric about $90^\circ$; all other observables are symmetric. Most of the existing
data are unfortunately converted into the forward hemisphere, so that obvious
violations of the symmetry are no longer visible, and artificial scatter might
lead to the false interpretation of higher partial waves.
Due to the
small values of the polarization observables, the fact that we calculate
them to only 3 significant places sometimes yields rough
lines in the figures and does not allow the finest details of the
calculations to be seen. 
Omitting $F$-waves in the full RRGM calculation yields 
at most unit changes in the last significant digit, which are too small to give
any noticable change in the figures. Therefore we do not generally compare
calculations with and without $F$-waves, but return to this point at the end of
the section.

The $R$-matrix analysis reproduces the rather old differential cross section 
data \cite{Brolley} well, considering the large error bars in fig. \ref{dddd-x}.
The RRGM calculations
do not reach the minimum at $90^\circ$ and overshoot the data below $60^\circ$,
with the full calculation being closer to the data.

\begin{figure}[h]
\centering
\includegraphics[width=8cm]{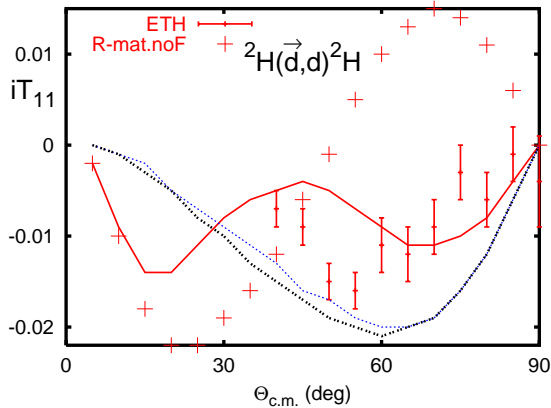}
\caption{(Color online) Vector analyzing power $iT_{11}$ for elastic deuteron-deuteron scattering
calculated at 3.0 MeV E$_{cm}$. The data are from the Z\"urich group 
\cite{Grueb-dd} at 3.0 MeV, transfered into the forward
hemisphere. The pluses are the $R$-matrix results omitting all $F$-wave contributions.}
\label{dddd-it11}
\end{figure}

In fig. \ref{dddd-it11} the rather small
vector analyzing-power data are compared to the calculations. All calculations
reproduce the small values with its relatively large errors.
The $R$-matrix analysis
shows structure that is not supported by the data. The effects of adding
TNF are negligible. Note the much smaller scale compared to \cite{HE4} here and in
the following analyzing powers.

\begin{figure}[h]
\centering
\includegraphics[width=8cm]{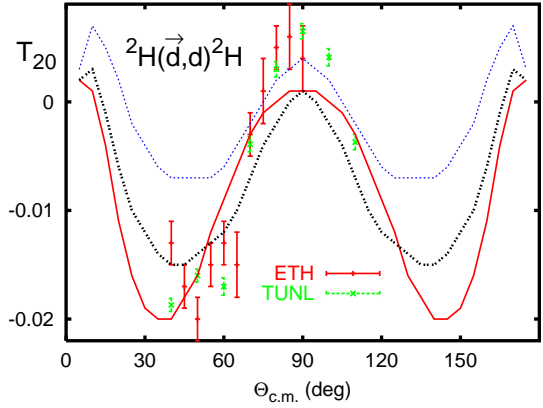}
\caption{(Color online) Same as fig. \ref{dddd-it11}, but for the tensor analyzing power
$T_{20}$. The data from the Z\"urich group \cite{Grueb-dd} are transfered into
the forward hemisphere, whereas the data of the TUNL group \cite{Crowe} are
shown as taken.}
\label{dddd-t20}
\end{figure}

The tensor analyzing power $T_{20}$ is one of the few variables with data also
in the backward hemisphere. Fig. \ref{dddd-t20} shows the data of the Z\"urich group
\cite{Grueb-dd}, with their scatter and large errors, and also the
data of the TUNL group
\cite{Crowe} with much smaller errors, covering the angular range up to 
$120^\circ$. The $R$-matrix analysis reproduces
the data nicely, not quite reaching the maximum at ninety degrees. The RRGM
calculation with AV18 alone misses the minimum around $45^\circ$ by about
a factor of two, but comes close to the maximal values. Adding TNF yields
relatively large effects, coming close to the $R$-matrix results and data.

\begin{figure}[h]
\centering
\includegraphics[width=8cm]{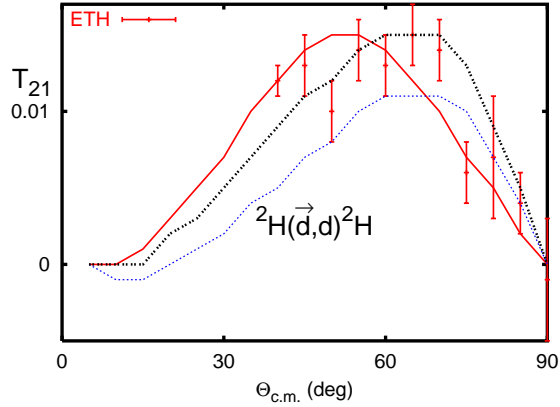}
\caption{(Color online) Same as fig. \ref{dddd-it11}, but for the tensor analyzing power
$T_{21}$.}
\label{dddd-t21}
\end{figure}

The $T_{21}$ data of the Z\"urich group \cite{Grueb-dd} show again considerable
scatter in both the values and uncertainties as seen in fig. \ref{dddd-t21}.
All calculations reproduce
these small polarizations reasonably well. The effects of TNF are again
small.

\begin{figure}[h]
\centering
\includegraphics[width=8cm]{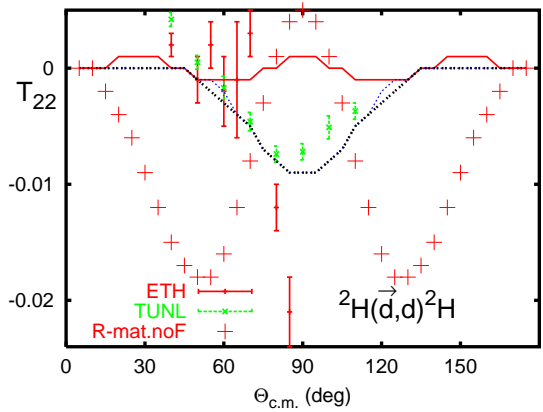}
\caption{(Color online) Same as fig. \ref{dddd-t20}, but for the tensor analyzing power
$T_{22}$. In addition the results of the $R$-matrix analysis omitting $F$-wave
contributions are shown as '+'.}
\label{dddd-t22}
\end{figure}

For the tensor analyzing power $T_{22}$ again exist data from the Z\"urich
\cite{Grueb-dd} and TUNL \cite{Crowe} groups. The Z\"urich data show a deep
minimum at $90^\circ$ with large relative errors, the TUNL data are about a
factor of two smaller in the minimum with the errors about the same factor
smaller, as seen in fig. \ref{dddd-t22}. 
The $R$-matrix analysis yields small positive and negative values with
no indication of the minimum. The two RRGM calculations are on top of each
other and reproduce the TUNL data nicely.

Let us now discuss the differences in the various calculations. As mentioned
before, the effects of $F$-waves in the RRGM calculations are negligible,
so we did not compare those cases.
As demonstrated in figs. \ref{dddd-it11}, \ref{dddd-t20}, \ref{dddd-t21}, 
and \ref{dddd-t22},
the effects of the TNF are small-to-negligible. For most observables,
the full calculations in the $R$-matrix or RRGM framework yield similar results,
especially if we take the smallness of the polarization values into account.
If we look at individual partial waves, however, this apparent agreement 
breaks down. Omitting the $F$-waves in the $R$-matrix analysis yields no effect in
the differential cross section, small changes for the $T_{21}$ angular 
distribution, a shallower minimum and maximum for $T_{20}$,
and large modifications
for the rest. The angular distribution of the vector analyzing powers 
$iT_{11}$ becomes $\sin (4 \Theta)$ with a minimal value of -0.022, as shown by
the pluses in fig. \ref{dddd-it11}, whose remnants are still visible in the
final result.
For $T_{22}$, the effects
of omitting the $F$-waves are displayed in fig. \ref{dddd-t22}. This suggests
that the $F$-waves are neccessary for some of the polarizations to compensate for
artifacts introduced by the lower partial waves. As mentioned in the
previous section for the $D$-wave and $P$-wave phase shifts and displayed in figs. 
\ref{D-wave-splitting}, \ref{dd_P}, and \ref{F-wave}
for $D$-, $P$-, and $F$-waves, the $R$-matrix analysis finds considerable J-splitting, 
whereas the RRGM
calculations do not support these findings. It is unclear at this point which
data cause this large J-splitting. The origin of this problem are the very few 
data in the backward hemisphere. Transforming the original backward hemisphere 
ETH-data into
the forward hemisphere might have introduced this artificial behaviour.

With the knowledge of the scattering and bound-state wave functions, various
radiative capture reactions can be calculated, at least in the long-wave-length
limit, for the complex RRGM wave functions used here.
For the reaction $^2$H(d,$\gamma)^4$He, new data at very low energies exist
from the TUNL group,
which are well reproduced, together with older data of this reaction by a
calculation similar to the one described above \cite{Trini}. Further work on the proton and
neutron capture reactions is under way \cite{Trini-dis}. Unfortunately for
these reactions, the data situation is very controversial; see the recent
calculation compared to data \cite{Giusi}.

\section*{Data needs and conclusion}

The $R$-matrix analysis uses many more data than discussed in the previous section
to determine the $R$-matrix parameters. Using these, it is possible to
interpolate or even extrapolate to energies where no data for the elastic
scattering or specific reaction are available, or to predict polarization observables
that have not been measured so far. We have chosen the energies
presented here by the requirement to have a maximum number of data sets at this
energy.

Compared to the previous calculations using the Bonn potential \cite{HE4},
the agreement between the parameter free RRGM calculation using the Argonne
v18 two-nucleon potential and the $R$-matrix analysis and data is much better.
The partial wave analysis allows to point to specific features that need
further studies, and additional or improved data. Due to the complexity of the
A=4 system, we cannot specify which part of the two- or three-nucleon
potential causes the differences seen in the previous sections. Therefore, we
can only point out which effective two-body interaction might be responsible.

Let us assume for the moment that the $R$-matrix is equivalent to the data; then
we can conclude from figs. \ref{0p-au}, \ref{1p-au}, and \ref{2p-dd-au}, that 
all the $S$-waves are very well reproduced and thus there is almost no room 
left for modifications of the central force. The $0^+$ triton-proton
phase shift below the neutron threshold might be the exception, which was used in
\cite{SCAT-LE}
together with the very low-energy data in $^3$He-neutron scattering and the
$^3$He(n,p){$^3$H} 
reaction, to advocate a slight reduction of the long-range and slight increase
of the short-range part of the central force.
The good agreement for the singlet $P$- and $D$-waves (see figs. \ref{1m-au} and
\ref{2p-au}, 
\ref{2p-dd-au} respectively), supports this further, whereas the $^1F_3$ triton-proton 
and $^3$He-neutron phase shifts disagree widely between $R$-matrix analysis
and RRGM calculation.

The deuteron-deuteron triplet $P$-waves indicate a rather strong spin-orbit
force in the $R$-matrix analysis, which is not quite met in the RRGM 
calculation, as seen in fig. \ref{dd_P}.
The [3 + 1] triplet $P$-waves show a smaller J-splitting in the $R$-matrix
analysis that the RRGM calculation follows nicely, without quite reaching the
values for the $^3P_1$ and $^3P_2$ phase shifts for the higher energies.
This mismatch also leads to differences in the
energy dependence, especially for the $0^-$ channel.

For the higher partial waves, which are small in the $R$-matrix analysis and the
RRGM calculation, relatively large differences occur. Whereas the RRGM calculation
reveals the dominance of the central force component by only a very weak
J-splitting in the elastic phase shifts, the $R$-matrix analysis yields
an appreciable splitting, the origin of which is still unclear.

Let us now compare directly to data. As mentioned in the previous section,
the [ 3 + 1 ] elastic scatterings and reactions are very sensitive to $P$-wave
matrix elements. The difference between the polarization of the heavy fragment
and the light one depends only on triplet-singlet transition matrix elements,
thus essentially singling out the $^3P_1 \longrightarrow {^1P_1}$ matrix element.
Since we realize that experiments with triton beams or targets are now 
problematic, we consider the reaction $^3$He(n,p){$^3$H} the perfect choice. Around
8 to 10 MeV neutron energy it would add to the already existing elastic scattering
data, and thus via unitarity, lead to a much more restrictive analysis. Taking
only the strongest transitions into account, $R$-matrix and RRGM calculations
yield widely different predictions for cross sections and analyzing powers.
Also, the TNF play a large role, due to the very large $0^-$ matrix element.
The cross section at forward angles differs by thirty percent.
Therefore, the energy region around the broad second $0^-$ resonance \cite{TILLEY-A4}
is of great interest.

For $d$ - $d$ elastic scattering, a few recent polarization data are available
\cite{Crowe} that contradict to a large extent the older ones. In order to
reduce the weight of the existing $iT_{11}$ and $T_{21}$ at higher energies,
new data would be highly welcome, preferably in the backward hemisphere.
Especially also cross section data with
smaller errors should improve the analysis and allow for a better comparison
between the various three-nucleon potentials.

The deuteron--deuteron fusion reactions are a very special case. Since there
the $F$-waves play a large role, they are essential in determining these
partial waves. Unfortunately the cross section data do not cover the very
forward and backward regions; hence, the $R$-matrix analysis is not forced to
reproduce these angle ranges. A few cross section measurements for both outgoing
channels would improve the situation tremendously. Also the dependence on the 
TNF is very strong at the extreme angles, as seen in figs. \ref{ddp-x}
and \ref{ddn-x}.

The polarization data of the proton channel are taken at almost twice as many
angles as are the neutron channel data. Since the relative errors are comparable,
the weight of the proton data is much higher in the analysis. The forward region
(for $T_{21}/A_{xz}$, the backward hemisphere) poses the main problem to
the $R$-matrix analysis. Especially for $iT_{11}$ and $T_{20}$,the differences between
the charge conjugate channels are large, which is not to be expected due to
charge symmetry. The differences due to the different thresholds are essentially
given by the variations of the RRGM calculations. Three-nucleon forces yield
large enough effects to make these polarizations a valuable tool to determine
TNF.

Before closing, we want to add a few remarks about the $A_y$ problem encountered
in deuteron-nucleon and proton-$^3$He elastic scattering and not seen here
in the $^4$He system. Since the mismatch between the maximal analyzing power of the
data and the calculations, for example using AV18 and UIX, is much larger for
the $p$ - $^3$He scattering than for deuteron-nucleon scattering, one might 
be tempted to conclude that the
origin of this deficiency is a missing $T=1$ force component. But then it should
also show up in the [ 3 + 1 ] channels considered here, which it does not.
This contradiction can be resolved by noting, that in the previous cases the
triplet $P$-waves matrix elements showed small J-spitting and had (essentially)
modulus one. Here, however, we find an appreciable J-splitting and due to
channel coupling, moduli much smaller than one, thus leading to a much more 
complex behavior of the analyzing powers.

We have shown in this paper how well an $R$-matrix analysis of the data can
represent them by a relatively small number of $R$-matrix parameters. In addition
to that, a parameter-free RRGM calculation allowed to show the agreement with
results from the Argonne v18 two-nucleon potential only and the effects of the
additional Urbana IX TNF. These effects are quite large for some partial 
waves or specific data. 
Therefore, we consider the $^4$He system a good place to study
the effects of TNF and to use the comparison with data in order to determine
the structure and radial dependence better. To learn more about the TNF,
other forces, like the Tucson-Melbourne force \cite{Sid} in its modified form
\cite{Sid-cor}, or one from
effective field theories \cite{van-Kolck} have to be applied also. 
Unfortunately, the RRGM calculations are very
time consuming, especially when the TNF is included.
Since every spin-isospin operator has to
be programmed  individually (and the radial dependence expanded in terms of
Gaussians), the recent Illinois force \cite{Steve} is currently
on the verge of feasibility \cite{Kirscher}.
The most accessible seems to be the Tucson-Melbourne force \cite{Sid-cor}, which 
contains no new operator compared to the Urbana IX force used here. We
plan a new calculation using this force.
The potentials derived from effective field theories can only be used when
a reliable configuration-space version is available.

Despite the above plea for additional and improved data, we consider the
whole set of data in the $^4$He system well suited to study the three-nucleon
forces. On the one hand there are clear-cut structures that vary only slowly
with energy, but in the reactions we found energy regions with a rapid change
in the observables, which could be used for a fine-tuning of the TNI. To be
not mislead by incorrect data, a procedure similar to the approach of
the Nijmegen-Group for the nucleon-nucleon data has to be carried out.
With the knowledge of the differences between the $R$-matrix analysis and the microscopic
calculation, we have started a new $R$-matrix analysis, where we use initially
only a few data sets, omitting all the suspicious ones with high $\chi^2$ values, 
and constraining the fit to give smaller J-splittings. Then we plan to add more
and more data sets, in order to learn which of them cause the J-splitting found so far.
The results of this tedious procedure will be published elsewhere.

Note added in proof: After completion of the manuscript, we became aware of the
work of A. C. Fonseca et al. \cite{Fonseca_NN, Fonseca_Delta}, in which they use
various NN-potentials, and come to similar results as we do for the AV18.
However, the many-nucleon force \cite{Fonseca_Delta} they include via
$\Delta$-excitation behaves differently from the UIX potential that we 
have used.

\begin{acknowledgements}
  This work is supported by the DFG under HO 780/9 and used resources at several
computer centers (RRZE Erlangen and HLRB M\"unchen).
We want to thank G.\ Wellein and
  G.\ Hager at the RRZE for their help.  The U.~S. Department of Energy
  supported the work of G.~M.~H. on this study.
\end{acknowledgements}

\bibliography{paper}

\end{document}